\documentclass{aa} 

\RequirePackage{longtable}

\usepackage{hyperref}
\usepackage{multirow}
\usepackage{longtable}
\usepackage{graphicx}
\usepackage{txfonts}

\usepackage[normalem]{ulem}
\usepackage{natbib}
\usepackage{color}

\begin{document}


\title{A photometric and spectroscopic study of eight semi-detached eclipsing binaries}
\titlerunning{Semi-detached eclipsing binaries}

\author{Yajuan Lei \inst{1,2,3},
        Guiping Zhou \inst{1,2,3},
        Liang Wang \inst{3,4,5},
        Guangwei Li \inst{6},
        Kai Li \inst{7},
        Tuan Yi \inst{8}
       }
\authorrunning{Yajuan Lei et al.}
\institute{
National Astronomical Observatories, Chinese Academy of Sciences, Beijing 100101, People's Republic of China \\
\email{yjlei@bao.ac.cn}
 \and
 State Key Laboratory of Solar Activity and Space Weather, Beijing 100190, China
\and
School of Astronomy and Space Science, University of Chinese Academy of Sciences, Beijing 100049, People's Republic of China
 \and
 Nanjing Institute of Astronomical Optics \& Technology, Chinese Academy of Sciences, Nanjing 210042, China;
          \and
 CAS Key Laboratory of Astronomical Optics \& Technology, Nanjing Institute of Astronomical Optics \& Technology, Nanjing 210042,China
          \and
 Key laboratory of Space Astronomy and Technology, National Astronomical Observatories, Chinese Academy of Sciences, Beijing 100101, China
          \and
Shandong Key Laboratory of Optical Astronomy and Solar-Terrestrial Environment, School of Space Science and Physics, Institute of Space Sciences, Shandong University, Weihai, Shandong, 264209, China
   \and
Department of Astronomy, School of Physics, Peking University, Yiheyuan Rd. 5, Haidian District, Beijing, 100871, People's Republic of China
 }

\date{Received ........; accepted ........}

\abstract
{ Semi-detached eclipsing binaries offer an exceptional opportunity to validate the evolutionary models of interacting binaries.
This entails determining the absolute parameters and evaluating the evolutionary status of the binary components through simultaneous analysis of both light and radial velocity curves. The Transiting Exoplanet Survey Satellite (TESS) provides high-precision light curves, while the Large Sky Area Multi-Object Fiber Spectroscopic Telescope Medium-Resolution spectroscopic Survey (LAMOST MRS) offers multiple epoch observations. The fusion of these data arises the chance to derive precise parameters for binaries.
}
{
The aim of this study is to estimate the absolute parameters of semi-detached binary components,
offer potential explanations for their evolutionary status, and investigate the long-term variations of orbital periods to explore their underlying causes.
}
{
By cross-matching the eclipsing binary catalog from TESS with
that from LAMOST MRS,
semi-detached eclipsing binaries with radial velocities coverage spanning more than 0.3 phases were authenticated.
The absolute parameters for these systems were determined by simultaneous modeling of
light curves and radial velocities using the Wilson-Devinney program.
Additionally, the secular orbital variations were further analyzed using O-C curves.
}
{
Eight semi-detached eclipsing binaries have been identified. Among them, seven feature primary stars situated within the main-sequence band,
while their secondaries are all in evolved stages. This suggests that these systems likely originated as detached binaries and have undergone a reversal of the mass ratio. However, TIC 428257299 is an exception where the primary is Roche lobe-filling, and its secondary has experienced mass loss events.
Additionally, TIC 8677671 and TIC 318217844 demonstrate secular cyclical changes of orbital periods.
Specifically, for TIC 8677671, the cyclical change could result from {\ magnetic activity or} a third body which is likely to be compact, with a mass of at least 2.97 M$_{\odot}$.
}
{}
\keywords{binaries: eclipsing - binaries: spectroscopic - stars: fundamental parameters} \maketitle

\section{introduction}
\label{sec:introduction}

Eclipsing binaries offer a preferred method for determining absolute parameters of each star independently of models.
This is achieved by concurrently analyzing light curves and variations in radial velocity (RV).
These stellar parameters play a crucial role in constraining both stellar structures and theoretical models of stellar evolution,
as demonstrated by previous studies \citep[e.g.,][]{Zha17, Par18, Soy20, Hel21, Li22}.

Eclipsing binary stars are categorized into three classes: detached, semi-detached,
and those in contact basis on their Roche lobe configurations  \citep{Kop59, Alc97}.
A semi-detached binary system involves one component filling its Roche lobe and transferring mass to its companion.
This interaction between binary components holds significant importance in stellar evolution
and serves as a means to validate corresponding models \citep{Iba06}.
Consequently, semi-detached binaries play a pivotal role in the study of various astrophysical phenomena,
such as mass transfer, accretion between components, and the loss of angular momentum through magnetic stellar winds.
Algol-type binary is generally semi-detached, comprising a main-sequence primary
ranging from mid-B to mid-F type and an F-G-K type giant or subgiant secondary that fills or overflows its Roche lobe \citep{Bud89}.
The two components may have comparable sizes in close proximity but significantly different masses.
Generally, the more massive primary is brighter than the less massive secondary.
During the formation of an Algol-type system, it is suggested that the initially massive component of a binary
will first fill the Roche lobe, subsequently transferring a substantial amount of mass to the other component.
This process leads to a reversal of the mass ratio and triggers a rapid evolution of the secondary \citep{Pol94, Li22}.

To gain a comprehensive understanding of interactive binary systems,
it is imperative to precisely determine the absolute parameters of both components within them \citep[e.g.,][]{Mal20}.
Although thousands of semi-detached binaries have been identified through light curves,
only a limited number have undergone accurate analysis, allowing for the simultaneous modeling of their light and RV curves.
Obtaining the absolute parameters for these systems proves challenging,
primarily due to the typically faint nature of their secondaries in the optical band.
\citet{Iba06} presented absolute parameters for 61 semi-detached Algol-type binaries,
while the well-known catalog by \citet{Bud04} listed 411 semi-detached Algol-type binaries.
However, a significant portion of these binaries is sourced from the catalogs of \citet{Bra80} and \citet{Sve90, Sve04},
introducing some uncertainty to the parameters as they were estimated solely based on light curves and empirical relationships.

In recent years, the increasing number of eclipsing binaries have been observed thanks to survey telescopes.
Sky surveys conducted by  the Convection, Rotation and planetary Transit \citep[CoRoT;][]{Auv09}
and Kepler \citep{Bor10} offered unparalleled precision and continuity.
These sustained, high-precision surveys play a pivotal role in determining the orbital parameters of binaries.
 The Transiting Exoplanet Survey Satellite \citep[TESS;][]{Ric15}, primarily aims at discovering transiting exoplanets,
provides continuous time-series photometry data.
Utilizing the Interlacing Code for Eclipsing binary Data validation,
Lightcurve Analysis Tool for Transiting Exoplanets, \citet{Mor21} reported thousands of TESS eclipsing binary candidates.

Fortunately, the second phase of LAMOST
initiated a  Medium-Resolution spectroscopic Survey (MRS; R $\sim$ 7500) in 2018.
This survey delivers medium-resolution spectra for hundreds of thousands of stars across multiple epochs, enabling the acquisition of time-series RVs \citep{Zon20}.
Moreover, LAMOST has conducted a low-resolution spectroscopic survey since 2012,
offering low-resolution spectral survey (LRS; R $\sim$ 1800 \AA) for over ten million stars \citep{Luo12}.
These spectra facilitate the determination of stellar atmosphere parameters.
By combining the high-precision photometry from TESS, time-series RVs from LAMOST MRS,
and stellar atmosphere parameters from LAMOST LRS, the orbital parameters of binaries
and the absolute parameters of both stellar components can be derived.

\par
By cross-matching the medium-resolution star catalog from LAMOST (DR9) with the TESS eclipsing binary catalog,
we initially identified a pool of candidate objects.
From this pool, we further selected samples for which the phase of the RV curve exceeded 0.3.
In the end, we successfully confirmed eight semi-detached eclipsing binary systems  according to LAMOST data,
including two double-lined eclipsing binaries (TIC 8677671 and TIC 318217844) and
six single-lined eclipsing binaries (TIC 312060302, TIC 428257299, TIC 168809669, TIC 406798603, TIC 158431889, and TIC 159386347).

\par
Both TIC 8677671 (WW Cnc) and TIC 318217844 (XZ UMa) were recognized as Algol-type eclipsing systems,
and their absolute parameters have been determined using the simple iterative method \citep{Sve90, Kim03, Bud04}.
Recently,  \citet{Lee24} presented the absolute properties of TIC 318217844 using data from TESS and the Bohyunsan Observatory Echelle Spectrograph.
They discovered high-frequency pulsation signals from its primary star,
which is a $\delta$ Sct pulsator, and classified the system as an oscillating eclipsing Algol-type binary.

TIC 312060302 (V1046 Cas) has been classified as an Algol-type eclipsing system by the All Sky Automated Survey for SuperNovae (ASAS-SN),
exhibiting an orbital period of 0.9805023 days \citep{Jay20}.
TIC 428257299 was identified as an Algol-type binary with an orbital period of 0.6389842 days according to ASAS-SN \citep{Jay20},
while it was classified as W UMa type with an orbital period of 0.6390052 days in the AAVSO International Variable Star Index (VSX) \citep{Wat06}.
TIC 168809669 (UW Boo) was proposed as an Algol-type eclipsing system binary with an orbital period of 1.00471 days in studies by \citet{Sve90, Wat06, Sam17} and \citet{Qia18}.
\citet{Che18} reported a slightly different period of 1.0046877 days based on Wide-field Infrared Survey Explorer (WISE) data.
The absolute parameters of  TIC 168809669 (UW Boo) were roughly estimated by \citet{Sve90, Kre04} and \citet{Soy06}.
Additionally, \citet{Man15} conducted an analysis of minimal times and derived absolute parameters based on its light curves.
TIC 406798603 (EW Boo) was classified as an Algol-type binary with an orbital period of 0.90636 days according to ASAS-SN data \citep{Jay20}.
A slightly different orbital period of 0.9086271 days was reported based on WISE data \citep{Che18}.
\citet{Zha15} determined the absolute parameters of TIC 406798603 solely from photometric data.
 \citet{Dou15} and \citet{Kim22} identified TIC 406798603 as a spectroscopic double-lined binary
and further refined its absolute parameters  through analyses of both photometric and spectroscopic data.
These studies explored the secular periodic variations of TIC 406798603.
Both \citet{Dou15} and \citet{Zha15} also furnished characteristic parameters of the pulsations of TIC 406798603,
categorizing the primary star as a $\delta$ Sct pulsator.
Additionally,  \citet{Kim22} identified additional pressure-mode pulsations,
confirming and integrating the frequencies discovered in the studies by \citet{Dou15} and \citet{Zha15}.

TIC 158431889  (V882 Lyr) has been recognized as a $\beta$ Lyrae-type eclipsing system,
displaying an orbital period of 0.5725256 days as indicated by ASAS-SN data \citep{Jay20}.
Additionally, \citet{Sla11} also classified it as a $\beta$ Lyrae-type based on Kepler observations.
Moreover, \citet{Lur17} classified TIC 158431889 (V882 Lyr) as exhibiting ellipsoidal variation.
In the case of  TIC 159386347, it was identified as an overcontact (W UMa-type) system based on Kepler data \citep{Sla11}.
The system parameters for both TIC 158431889 and TIC 159386347 were established using an artificial intelligence method \citep{Sla11, Arm14}.
However, as per the available information, the absolute parameters for these two systems have not yet been disclosed.

In this paper, based on simultaneously modeling light curves from TESS and RVs from LAMOST MRS,
we performed calculations and analyses of the system's parameters and absolute stellar parameters for all eight semi-detached binary systems with the assistance of the Wilson-Devinney (W-D) program.
The organization of this paper is as follows: In Sect. 2, we introduce the photometric and spectroscopic data utilized in our analysis.
Section 3 presents the results obtained by modeling the light and RV curves using the W-D program.
Finally, discussions and conclusions are provided in Sects. 4 and 5, respectively.

\section{Data}
\label{sec:data}

\begin{table*}
\tiny
\vspace{0.9cm}
\setlength{\tabcolsep}{2.pt}
\caption{Basic information for the eight systems studied in the work.}
\label{table:tb1}
\centering
\begin{tabular}{l c c c c c c c c}
\hline\hline
Objects                      & TIC 0312060302                   & TIC 428257299                  & TIC 8677671           & TIC 318217844                    & TIC 168809669               & TIC  406798603                & TIC 158431889                    & TIC 159386347           \\
\hline
 R.A. (J2000)    & $00^{h}40^{m}44.2^{s}$        &$03^{h}40^{m}11.0^{s}$         & $09^{h}09^{m}48.6^{s}$         & $09^{h}31^{m}24.7^{s}$         &  $14^{h}20^{m}59.6^{s}$        &  $15^{h}02^{m}46.1^{s}$       & $19^{h}09^{m}33.9^{s}$        & $19^{h}21^{m}23.2^{s}$        \\
 Dec. (J2000)   & $+58^{\circ}50\arcmin53.8\farcs$ &$+51^{\circ}54\arcmin42.9\farcs$ & $+30^{\circ}25\arcmin36.8\farcs$ & $+49^{\circ}28\arcmin04.4\farcs$ & $+47^{\circ}06\arcmin45.1\farcs$ &$+37^{\circ}54\arcmin36.4\farcs$ & $+43^{\circ}05\arcmin55.4\farcs$ & $+44^{\circ}49\arcmin03.0\farcs$ \\
 Tycho  ID                   &   3667-826-1                 &  ...                      &   2492-824-1       &  3429-1530-1               &    3472-246-1           &    3044-443-1            &  ...                 &   3146-286-1                 \\
 GSC    ID                   &   03667-00826                &  ...                           &  	 02492-00824     &  03429-01530                  &      ...                         &    03044-00443            &  ...                 &   ...                           \\
 GCVS                        &    V1046 Cas                  &  ...                           &  WW Cnc             &   XZ UMa                         &     UW Boo                       &  	  EW Boo                  & V882 Lyr                 &  ....                          \\
 TESS (mag)                  &   11.39                       &  	11.60                     &  10.02              & 10.00                            &      10.32                       &     10.13                     &  12.00             &  11.44                          \\
 $P_{orb}$/TESS (days)     &   0.980500                     &   0.638983                    &   1.115977           & 1.222293                      &   1.004712                     &  0.906341                  & 0.572531            &  0.807987                   \\
 $P_{orb}$/O-C (days)      &   0.980499                     &   ...                          &   1.115957            & 1.222310                     &   1.004711                      &  0.906349                     & ...             & 0.808081                         \\
\hline
\end{tabular}
\footnotesize {}
\end{table*}

\begin{table}
\caption{Used data of the eight  binaries from TESS.}
\label{table:tb2}
\centering
\begin{tabular}{l c}
\hline\hline
Objects            &  Sectors            \\
\hline
TIC 0312060302     & 17, 18, 58          \\
TIC 428257299      & 18, 19, 59          \\
TIC 8677671        & 21, 48              \\
TIC 318217844      &  21                 \\
TIC 168809669      & 16, 23, 49, 50      \\
TIC 406798603      & 23, 24, 50, 51      \\
TIC 158431889      & 40, 41, 53, 54, 55  \\
TIC 159386347      & 14, 15, 40, 41, 55  \\
\hline
\end{tabular}
\footnotesize {}
\end{table}

\subsection{Photometry and orbital periods }

TESS was launched on April 18, 2018, equipped with four identical and wide-field cameras designed to monitor a strip of the sky in a $24^{\circ}\times 90^{\circ}$ configuration. The passband of TESS covers the range from 600 to 1000 nm.
Employing each of its cameras, TESS systematically observes each sky strip for at least 27 days and nights.
During the primary mission of TESS, Full Frame Images (FFIs) were taken every 30 minutes.
In the first mission extension, FFIs were taken every 10 minutes.
Starting from Cycle 5, the second mission extension, FFIs will be taken every 200 seconds.
The Target Pixel Files (TPFs) of TESS represent the most raw form of target-specific data available at the Mikulski Archive for Space Telescopes (MAST).
 There is also 2-minute cadence during the primary mission.
There are two types of TPFs: 2-minute cadence (available for all mission cycles) and 20-second cadence (available from Cycle 3 onwards).
For the purposes of this study, the data with 2-minute cadence were utilized to  avoid overlap.
The TESS satellite boasts a pixel scale of 21 arcseconds, implying a higher likelihood of crowding in the observed field.
In each sector, an aperture of approximately 3 by 3 pixels with slight variations is employed.
An important consideration is the potential contamination of the aperture by sources located beyond a radius of 63 arcseconds from the target.
To address this concern, we conducted a thorough examination of the vicinity of the target sources to identify any additional stars
using Gaia Data Release 3 \citep[DR3;][]{Gaia16, Gaia23}.
If other stars were found, their Gaia magnitudes will be converted to TESS magnitudes to calculate the contribution of contaminated light.
This conversion was performed based on the method outlined by \citet{Sta19}.

\begin{table}
\tiny
\caption{Contaminated sources within 63 arcseconds for the eight  binaries.}
\label{table:tb3}
\centering
\begin{tabular}{|l c | c c c|}
\hline\hline
Objects          & $G_{mag}$  & r($^{\prime \prime}$)& Gaia Sources ID      &   $G_{mag}$  \\
\hline
 \multirow{12}{*}{TIC 0312060302}  & \multirow{12}{*}{11.69}    &      1.4        &425352056105918080  &   13.64  \\
                 &          &     24.5       &425352056103563392  &   14.68 \\
                 &          &     28.1       &425352051800819328  &   13.09 \\
                 &          &     33.3       &425352051800819456  &   11.97 \\
                 &          &     42.2       &425351953024354432  &   13.05 \\
                 &          &     45.2       &425351953024354432  &   15.11 \\
                 &          &     47.4       &425351953024354432  &   15.91 \\
                 &          &     47.9       &425351953024354432  &   16.36 \\
                 &          &     51.8       &425351953024354432  &   17.04 \\
                 &          &     55.6       &425351953024354432  &   16.23 \\
                 &          &     57.9       &425351953024354432  &   14.56 \\
                 &          &     62.6       &425351953024354432  &   15.88 \\
 \hline
 TIC 428257299   & 12.22    &     49.4       &443682632922733952  &  16.66   \\
 \hline
 TIC 8677671     & 10.36    &     8.9        &699818436453891072  & 14.44   \\
 \hline
 TIC 318217844   & 10.24    &     ...        & ...                & ...      \\
 \hline
 TIC 168809669   & 10.56    &    45.7        &1506949959396228736 & 16.66   \\
 \hline
 \multirow{2}{*}{TIC  406798603}  & \multirow{2}{*}{10.26}    &    1.7        & 1295253976312344704 & 14.46   \\
                 &          &    42.7       & 1295253976313611648 & 17.04   \\
 \hline
 \multirow{2}{*}{TIC 158431889}   & \multirow{2}{*}{12.25}    &      24.3     & 2102608322164332032  & 17.41  \\
                 &          &      57.0     & 2102607978566958976  & 17.21  \\
 \hline
  \multirow{8}{*}{TIC 159386347}   & \multirow{8}{*}{11.72}    &      23.6     & 2126998646862919936  & 12.38   \\
                 &          &      27.9     & 2126998651164072704  & 16.40   \\
                 &          &      32.5     & 2126998719883551744  & 15.87   \\
                 &          &      33.7     & 2126998685523824640  & 17.48   \\
                 &          &      45.9     & 2126998719883548032  & 17.05   \\
                 &          &      47.3     & 2126995730585767680  & 13.31   \\
                 &          &      52.2     & 2126997173694759040  & 16.87   \\
                 &          &      53.5     & 2126998788603612288  & 14.95   \\
\hline
\end{tabular}
\footnotesize {The contaminated sources presented in the table have a magnitude difference with the target star within 6 magnitudes.}
\end{table}

For TESS observations of heavily contaminated sources, alternative datasets from Kepler, ASAS-SN, and Gaia were employed.
The Kepler mission's spacecraft is equipped with a wide-angle telescope featuring a 1 m diameter mirror to monitor a field of view (FOV) of $16.1^{\circ}$,
covering a sky area of 115.6 square degrees.
The passband of Kepler ranges from 420 to 900 nm and light curves with a 30-minute cadence are available for most stars.
ASAS-SN, monitoring the entire visible sky down to V $\sim$ 17 mag, provides light curves in V-band and
g-band magnitudes with a cadence of approximately 1-3 days \citep{Jay19}. Although the Gaia band observations are sparse compared to the TESS data,
they offer high precision in position.
The median position uncertainties are 0.01-0.02 mas for G $<$ 15 and 1.0 mas at G = 21 mag.
Consequently, the G-band photometry from Gaia was used as a supplementary source for systems that experienced significant contamination in the TESS data.

The basic information of the eight systems is shown in Table \ref{table:tb1}.
The used light curves were downloaded from MAST database.
The Science Processing Operations Center (SPOC) pipeline provides two fluxes for each light curve:
the Simple Aperture Photometry (SAP) flux and the Pre-search Data Conditioning SAP (PDCSAP) flux \citep{Jen10, Jen16}.
In our analysis, the PDCSAP fluxes were used because their long-term trends were eliminated.
The data from TESS used in this work are shown in Table \ref{table:tb2}.
Table \ref{table:tb3} shows these targets which have contaminated the TESS light curves of the studied eight systems.

The orbital periods of these systems have been calculated by some previous studies.
To further refine their orbital periods and investigate secular variations,
 we scrutinized the `O - C Gateway' database\footnote{\url{http://var2.astro.cz/ocgate/}} \citep{Pas06},
which has been scanned over an extended period, spanning even more than a century.
TIC 428257299, TIC 158431889, and TIC 159386347 are absent from the databases, whereas the remaining five systems are included.
Due to the limited number of occurrences of shallow secondary eclipses in all the systems,
we opted to utilize primary timings for the analysis.
Subsequently, the available primary minima data for the systems, including visual, photographic, photoelectric CCD, etc.,
obtained from the `O - C Gateway', were prepared for analysis.
Additionally, primary minima times from TESS data were incorporated into the analysis of the periods for these systems.
The conversion between Heliocentric Julian Day (HJD) and Barycentric Julian Day (BJD) timings was conducted using online applets \citep{Eas10}
\footnote{\url{http://astroutils.astronomy.ohio-state.edu/time/}}.
The results indicate that the orbital periods of TIC 8677671, TIC 318217844, and TIC 406798603 exhibit secular variations,
and TIC 0312060302 and TIC 168809669 do not display obvious secular variations in their orbital periods.
Considering that TIC 406798603 has been previously examined in the study by \citet{Kim22},
our analysis concentrated exclusively on the O-C curves of TIC 8677671 and TIC 318217844.

The equation of  a quadratic plus the light-time effect (LITE) by a third body is usually used for the analysis of changes in orbital periods:
\vspace*{-0.1 cm}
\begin{equation}
T =  T_0 + E \times P + Q \times E^2 + {{a_{12}sini}\over{c}}[{{1-e^2}\over{1+ecos\nu}} +esin\omega ],
\end{equation}

where $T_0$ is reference time for primary eclipse, $P$ stands for the orbital period, $E$ and $Q$ represent the epoch number,
the variation in the period rate, respectively.
The final term in the equation accounts for the time delay resulting from the motion of
the eclipsing binary around the common center of mass with the unseen star in the close binary,
and $a_{12}$, $e$, $i$, $\omega$ and $\nu$ are the orbital parameters  \citep{Irw59}.

The O - C analysis of TIC 318217844 suggests a secular
orbital period decrease of the binary system. The rate of period
decrease turns out to be dP/dE = 3.7 $\times$ $10^{-10}$ days $cycle^{-1}$ or
dP/dt = 1.1 $\times$ $10^{-7}$ days $yr^{-1}$.
Additionally, it is seen that there are  cyclic variations in the O-C graphs of the two systems, respectively (Fig. 1).
To account for this cyclic variation, the theoretical fit was applied to all minima data.
The orbital parameters of the third body are detailed in Table \ref{table:tb4}.
TIC 8677671 also shows the fourth body whose parameters are also shown in Table \ref{table:tb4}.
TIC 159386347 is not included in the `O-C Gateway', but was observed by both Kepler and TESS.
We analyzed its O-C curve based on data from Kepler (Quarters 0-17) and TESS, and found no secular trend.
The periods obtained based on O-C method are in good agreement with those calculated by the data of ASAS-SN or TESS \citep{Jay20, Mor21}.
The periods are listed in Table \ref{table:tb1}.

\begin{figure*}
\begin{center}
\includegraphics[width=14cm,height=7cm,angle=0,clip]{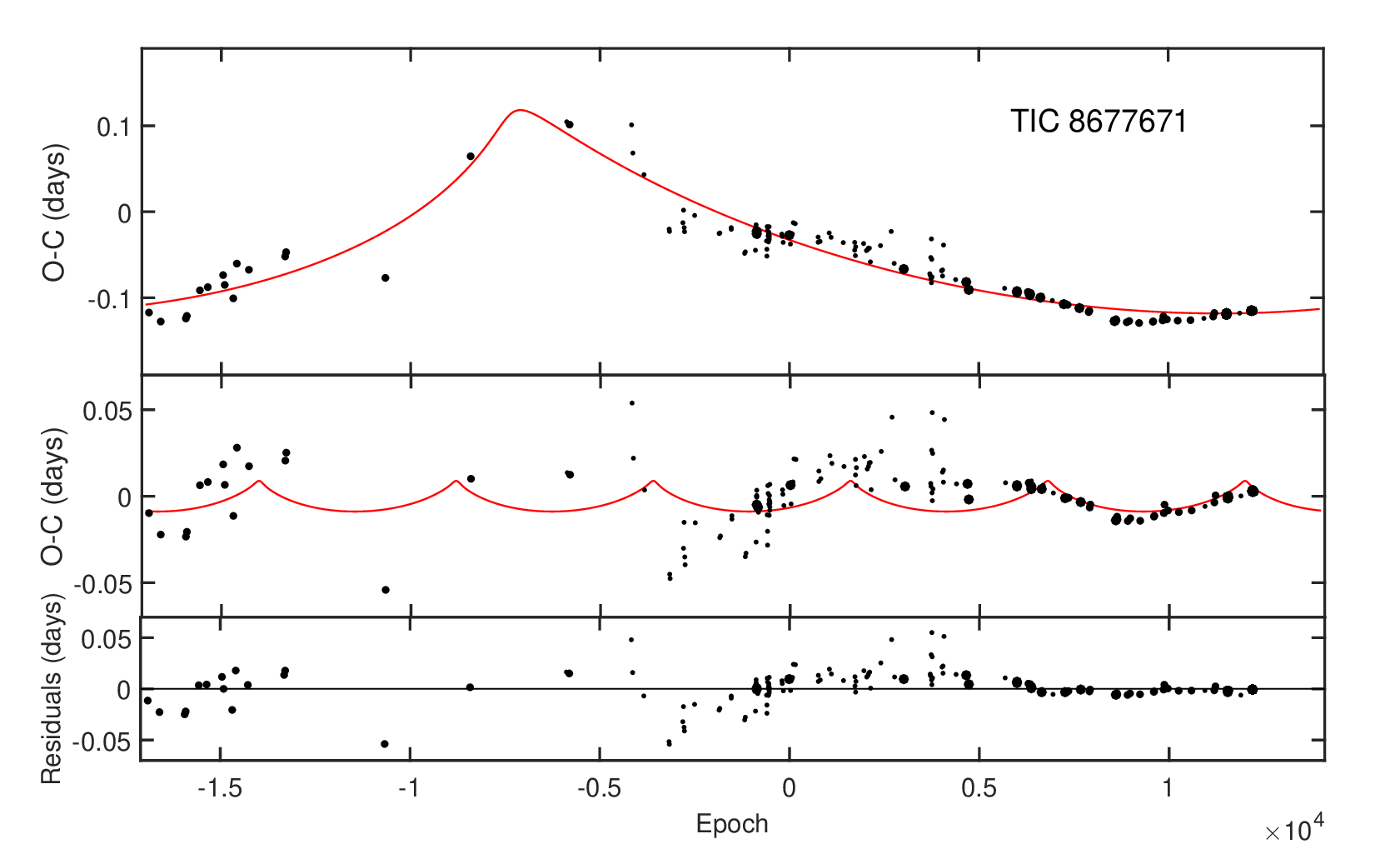}
\includegraphics[width=14cm,height=7cm,angle=0,clip]{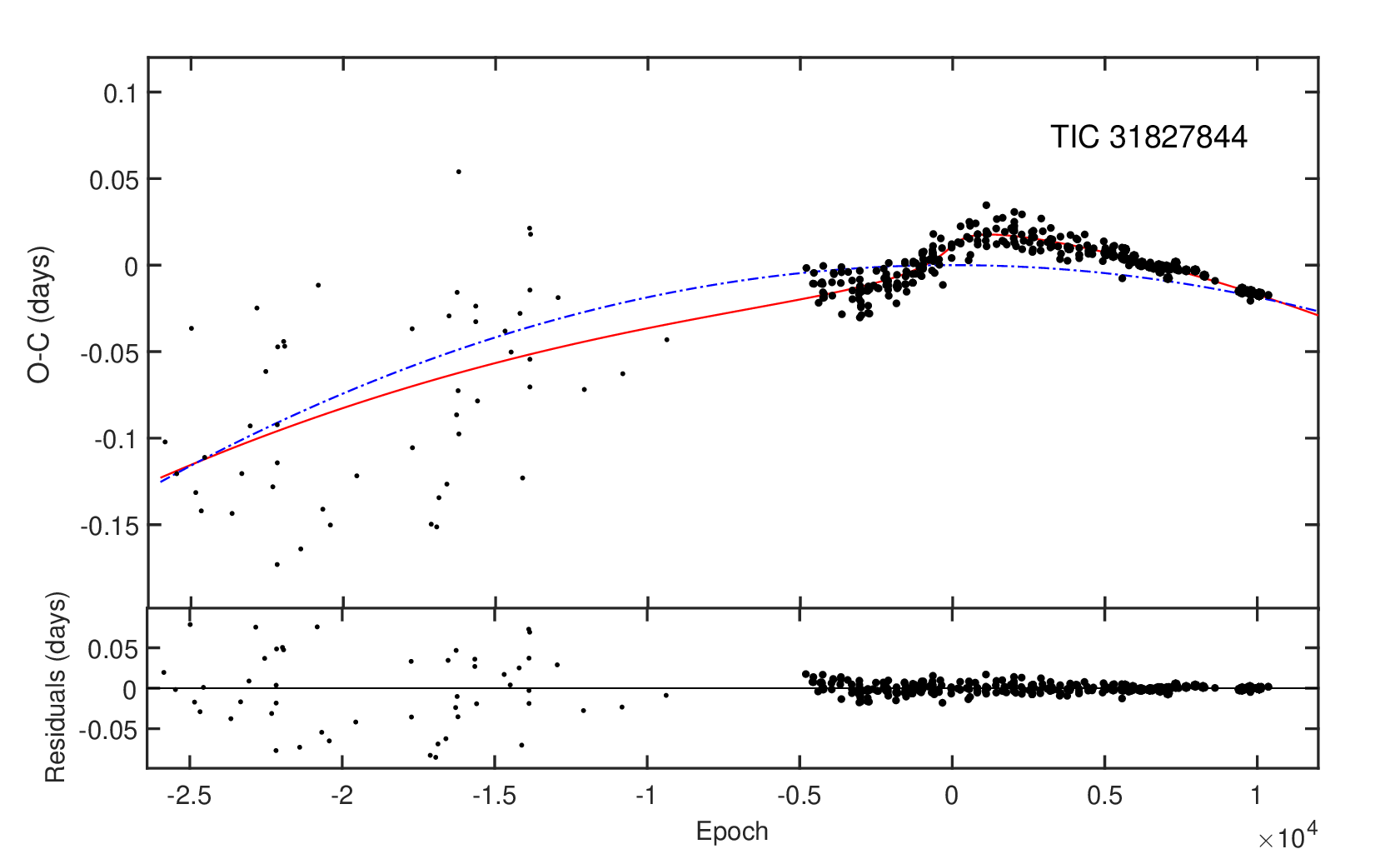}
\caption{O-C diagrams of TIC 8677671 and TIC 318217844 with LITE ephemeris.}
\end{center}\label{fig:fig1}
\end{figure*}

\begin{table}
\vspace{0.5cm}
\tiny
\caption{Parameters  calculated as a result of O-C analyzes.}
\label{table:tb4}
\centering
\begin{tabular}{l c c}
\hline\hline
\renewcommand{\tabcolsep}{0.10pc}
Objects                               & TIC 8677671          & TIC 318217844       \\
\hline
$JD_{0}$ (BJD)                        & 2446005.6365(56)     &  2446592.5617(15)    \\
P (days)                              & 1.1159629(7)         &  1.2223071(20)      \\
dP/dE($10^{-10}$days $cycle^{-1}$)    & ...                  & -1.8538(4)    \\
$T_3$ (BJD)                           & 2437775(400)         &  2446625(561)       \\
$P_3$ (yr)                            & 97.9(3.9)            &  118.9(1.0)         \\
$a_{12}$$sini_3$ (AU)                 & 22.27(36)            &  4.70(46)          \\
$A_3$ (days)                          & 0.1182(19)           &  0.0180(18)         \\
$\omega_3$($^{\circ}$)                & 61.7(6.0)            &  25.4(12.2)         \\
$e_3$                                 & 0.83(5)              &  0.83(13)           \\
f($m_3$)(M$_{\odot}$)                 & 1.15(16)             &  0.0074(1)          \\
$m_3$ (M$_{\odot}$) (for i = 90 deg)  & 2.97(23)             &  0.4225(23)         \\
$T_4$ (BJD)                           &  2494231(119)        & ...  \\
$P_4$ (yr)                            &  15.89(6)            & ...  \\
$a_{12}$$sini_4$ (AU)                 &  1.54(32)            & ...  \\
$A_4$ (days)                          & 0.0089(18)           & ...   \\
$\omega_4$($^{\circ}$)                &  95(50)              & ...  \\
$e_4$                                 & 0.83(28)             & ...  \\
f($m_4$)(M$_{\odot}$)                 &   0.0145(20)         & ...  \\
$m_4$ (M$_{\odot}$) (for i = 90 deg)  &   0.41(3)            & ...  \\
\hline
\end{tabular}
\footnotesize {}
\end{table}

\subsection{Spectroscopy}
\label{section:2.2}

LAMOST is a quasi-meridian reflecting Schmidt telescope with an effective aperture of approximately 4 m,
and a field of view of $5^{\circ}$ in diameter  \citep{Cui12}.
It can simultaneously observe 4000 targets with wavelengths ranging from 3800 to 9000 \AA.
During the first regular survey (2012$-$2017), LAMOST operated in a low-resolution mode (R $\sim$ 1800 \AA) \citep{Luo12}.
In September 2018, LAMOST initiated the MRS, with each exposure lasting 10 or 20 minutes.
MRS utilizes almost half of the telescope's observation time to enable time-domain astronomy,
employing both blue (4950 - 5350 \AA) and red (6300 - 6800 \AA) arms \citep{Liu19}.
The LAMOST MRS aims to observe approximately 200,000 stars, averaging 60 observations per star, from 2018 to 2023.
As a result, time-series RVs of a large number of objects can be obtained.

 The RVs of the eight systems observed by LAMOST MRS were computed utilizing cross-correlation functions (CCFs) \citep{Mer17},
and the RV measurements are provided in Table \ref{table:A1} and Table \ref{table:A2}.
The spectra from the Phoenix library serve as the templates (Husser et al. 2013).
We used the parameters (effective temperature $T_{eff}$, surface gravity log $g$,  metallicity [Fe/H]) from TIC v8.2 as reference parameters.
For each sample, the parameters of the templates were determined by selecting the closest match to the values provided by TIC v8.2 \citep{Pae21}.
In cases where [Fe/H] was not provided, it was set to 0.0.
During the calculation process, we tried to use both red and blue spectra, respectively. Overall, the results from the blue arm were more pronounced, so we opted to use the blue arm.

In addition to the RVs calculated from MRS,
the orbital solution for binary stars also requires atmospheric parameters derived from low-resolution spectra.
Utilizing the Universite de Lyon Spectroscopic analysis Software (ULySS),
constructed upon the ELODIE stellar library \citep{Pru01, Pru07, Kol09, Wu11, Wu14},
the LAMOST pipeline furnishes atmosphere parameters ($T_{\rm eff}$, log $g$,
and [Fe/H]) for stars ranging from later A to K types observed by LAMOST LRS.
While the LAMOST pipeline employs a single-star model. To estimate the effective temperature of each target, we applied the broadband Spectral Energy Distribution (SED) fitting method as described by  \citet{Yi2022}.  A comprehensive description can be found in \citet{Yi2022}. Initially, we gathered archival broadband photometry from various surveys, including 2MASS \citep[J, H, K bands]{Skr06}, WISE \citep[W1-W4 bands]{Wri10}, Gaia \citep[Bp, Rp, Gp bands]{Gaia23}, and Galex \citep[NUV and FUV bands]{Mar05}. We also incorporated trigonometric parallax-derived distances $D$ and interstellar extinction ($A_\mathrm{G}$) from Gaia DR3 \citep{Gaia23}. The model SEDs are Kurucz synthetic photometry models obtained from the VOSA website \footnote{\url{http://svo2.cab.inta-csic.es/theory/newov2/syph.php}}\citep{Bayo08}.
$T_{eff}$ represents the effective temperature of the binary system derived from the SED analysis.
 The results of SED analysis are shown in Fig. 2.
After acquiring the temperature ratio $T_2/T_1$ and radius ratio k($R_2/R_1$), we derived each component's effective temperature using
Equation (3) given by \citet{Zwi03}:
\begin{equation}
T_1 =  (((1 + k^2)T_{eff}^4 / (1+k^2(T_2/T_1)^4))^{0.25}.
\end{equation}

In addition, we compare $T_{eff}$ values obtained from SED analysis with those from LAMOST in Table 5 to evaluate
how accurately LAMOST estimates $T_{eff}$ values for eclipsing binaries.
In the Table 5, the errors of $T_{eff}$ values from LAMOST are assessed as the variance among all available temperatures obtained from LAMOST LRS.
In cases where only one temperature is available for a binary system,
the temperature error is assumed to be the value provided by the LAMOST pipeline.
For TIC 0312060302, whose spectrum is considered to correspond to a hot star, the effective temperature of 10,078 K,
as reported by \citet{Xia22} who calculated it using LAMOST observations, is shown in Table 5.
The results indicate that the $T_{eff}$ values from the two methods are consistent for our samples where the light from secondaries contributes less than 25\% of the total light (refer to the results in Tables 6 and 7 below).

\begin{figure*}
\begin{center}
\includegraphics[width=8.7cm,height=5.7cm,angle=0,clip]{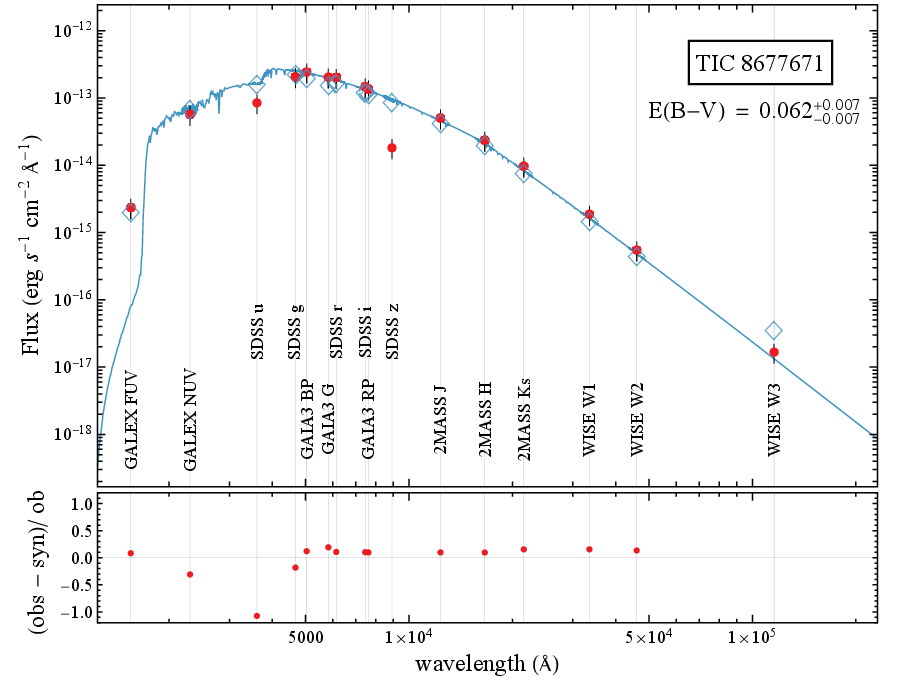}
\includegraphics[width=8.7cm,height=5.7cm,angle=0,clip]{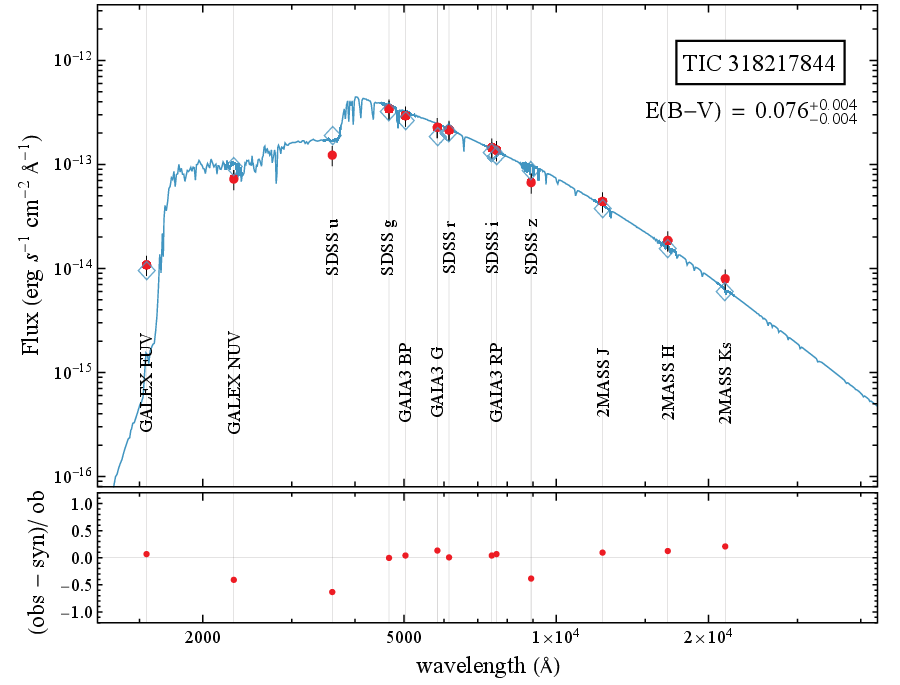}
\includegraphics[width=8.7cm,height=5.7cm,angle=0,clip]{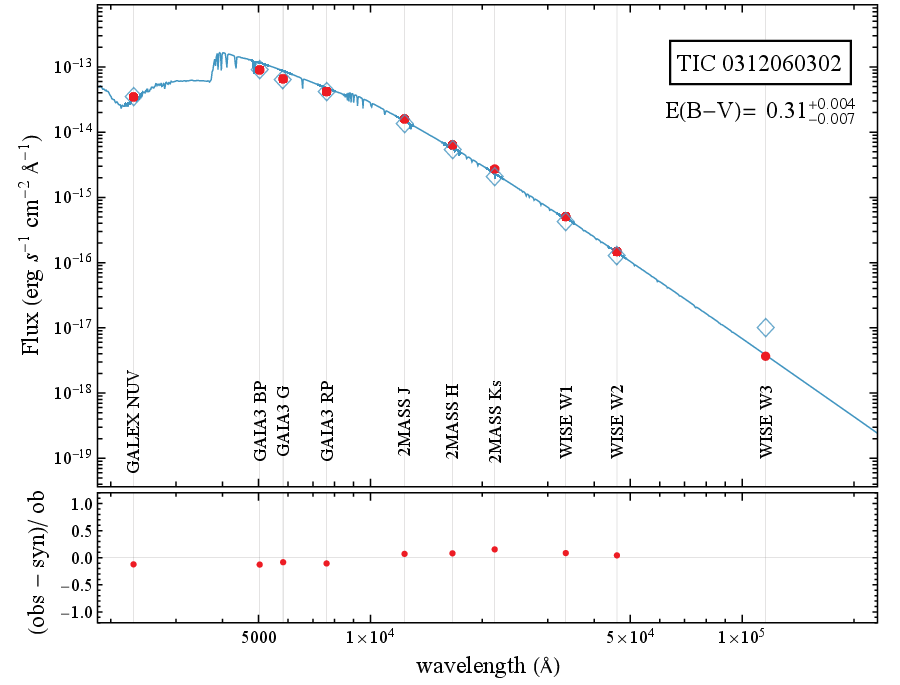}
\includegraphics[width=8.7cm,height=5.7cm,angle=0,clip]{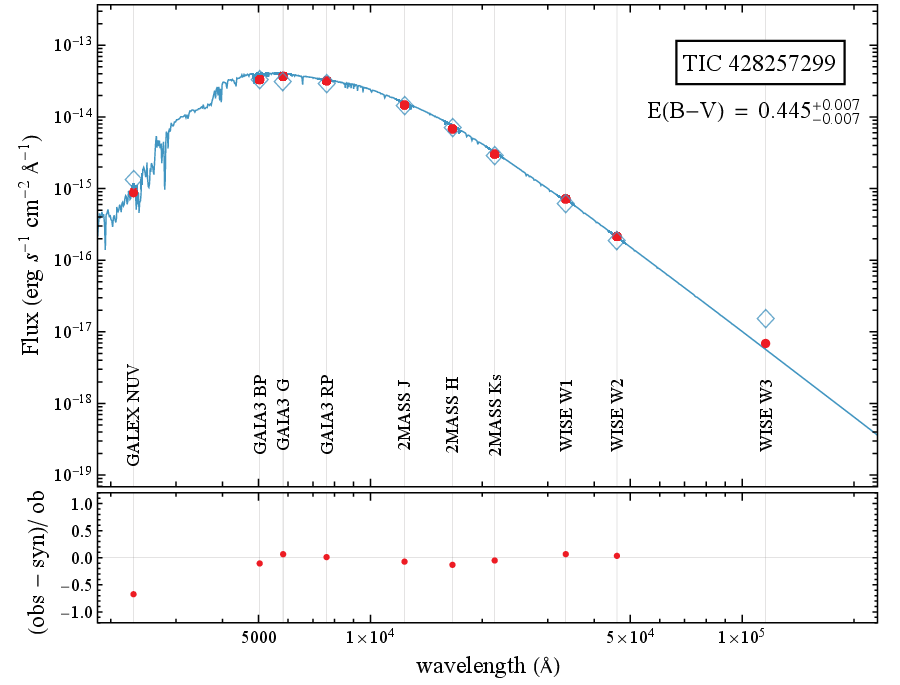}
\includegraphics[width=8.7cm,height=5.7cm,angle=0,clip]{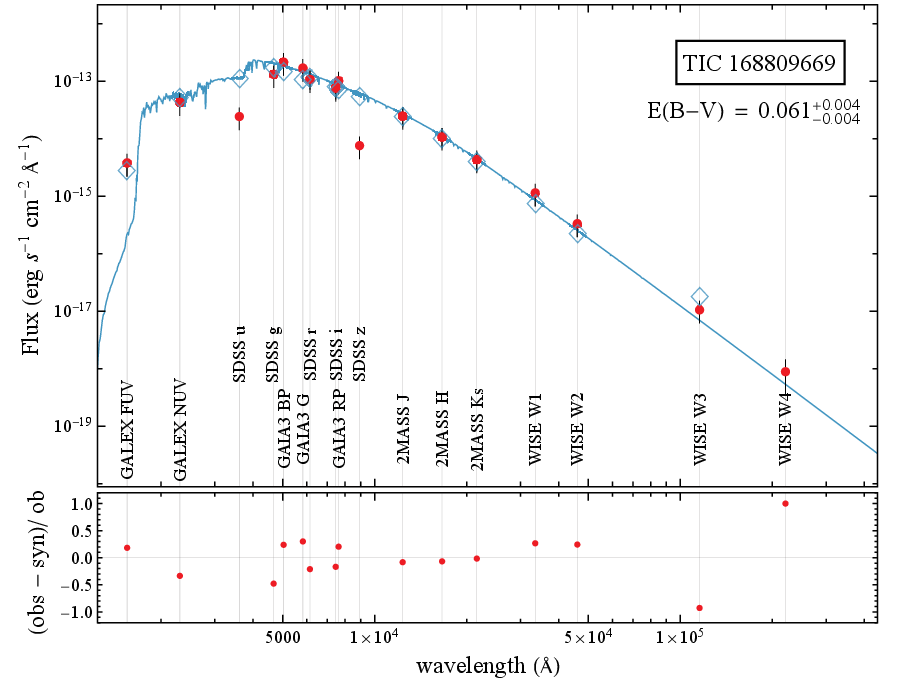}
\includegraphics[width=8.7cm,height=5.7cm,angle=0,clip]{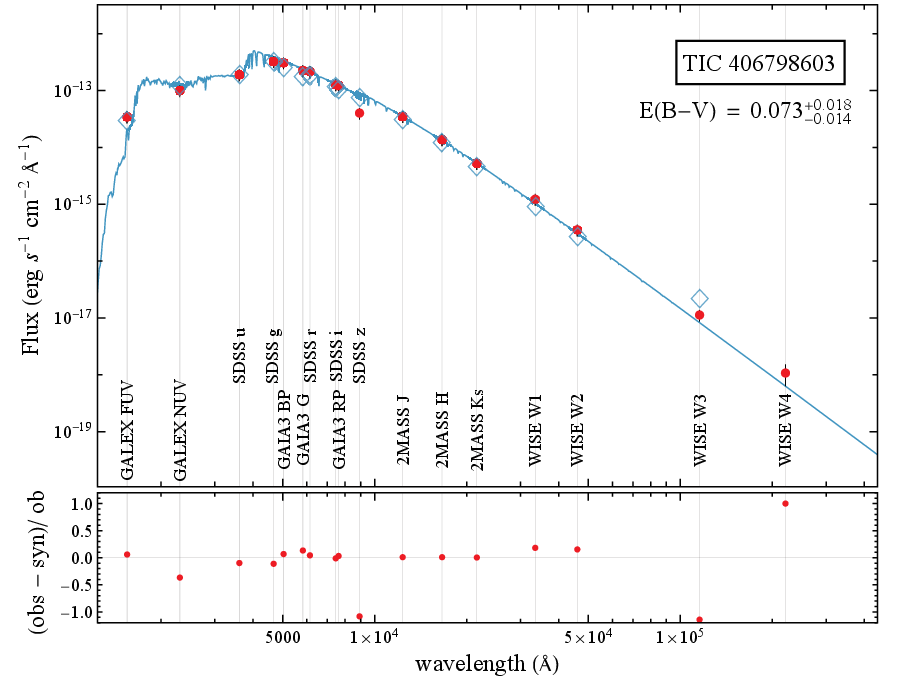}
\includegraphics[width=8.7cm,height=5.7cm,angle=0,clip]{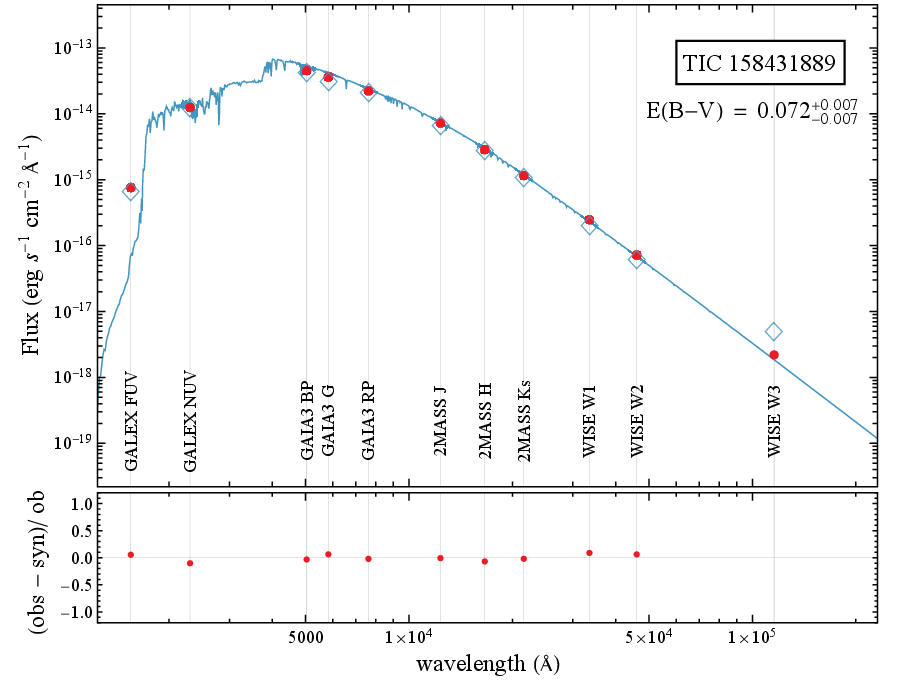}
\includegraphics[width=8.7cm,height=5.7cm,angle=0,clip]{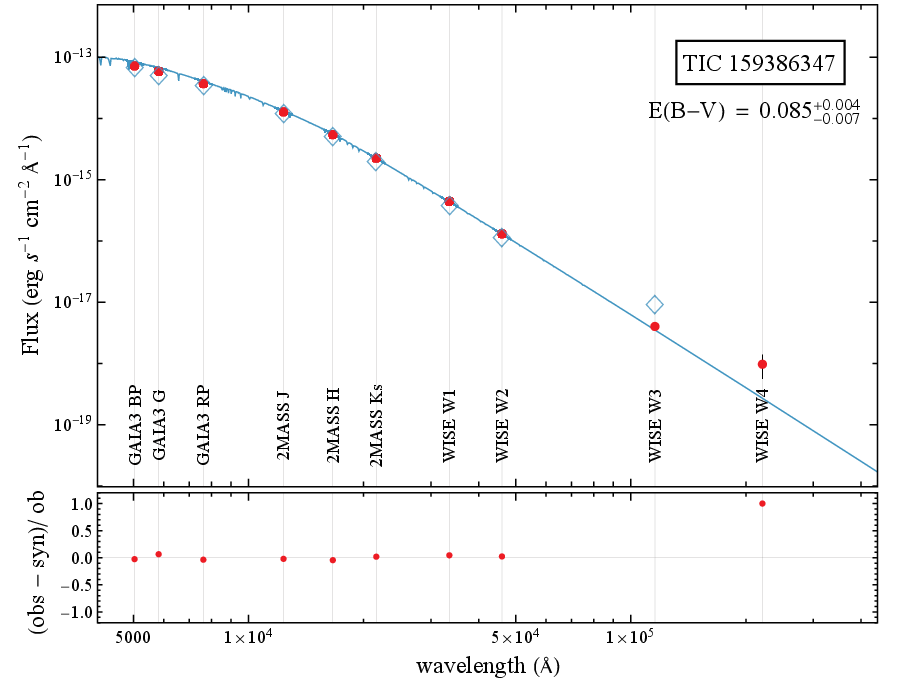}
\caption{ Single-model fit for the broadband SEDs of the eight targets.
The red dots represent photometric fluxes from various bands, as indicated and labeled by the vertical gray lines.
The blue curve shows the best-fit Kurucz SED with interstellar dust extinction corrected.
The blue diamonds represent the best-fit synthetic photometric fluxes.
The lower sub-panel shows the relative residuals, defined as (observed flux - synthetic flux) / observed flux.}
\end{center}\label{fig:fig2}
\end{figure*}

\begin{table*}
\caption{Effective temperatures from SEDs and LAMOST, respectively.}
\label{table:tb5}
\centering
\begin{tabular}{l c c c}
\hline
\hline
    sources      &  $T_{SED}$         &  $T_{LAMOST}$          \\
\hline
TIC 8677671      & 6517 $\pm$ 140     &  6710 $\pm$ 68         \\
TIC 318217844    & 7621 $\pm$ 143     & 7620 $\pm$ 113         \\
TIC 0312060302   & 10256 $\pm$ 261    & 10078 $\pm$ 141        \\
TIC 428257299    &  6881 $\pm$ 163    & ...         \\
TIC 168809669    &  7167 $\pm$ 162    & 7282 $\pm$ 12          \\
TIC 406798603    &  8281 $\pm$ 158    & 8299 $\pm$ 35          \\
TIC 158431889    & 7378 $\pm$ 111     & 7047 $\pm$ 18          \\
TIC 159386347    &  7091 $\pm$ 117    &  7021 $\pm$ 58         \\
\hline
\end{tabular}
\end{table*}

\section{Orbital Solutions}
\label{section:3}

The W-D program is a sophisticated tool for analyzing binary systems, incorporating principles of physics and numerical convergence.
It utilizes Roche geometry to depict the surface profiles of stars in binary systems and employs light and RV data to infer binary parameters \citep{Wil71, Wil79, Wil14}.
In this investigation, we utilized the 2015 version of the W-D program to evaluate both the system and absolute stellar parameters of eight eclipsing binaries.
The light curves analyzed by the W-D program accounted for identified contamination from external sources.

The $T_1$, as discussed in Sect.~\ref{section:2.2}, was served as input for the primary temperatures in the W-D program.
Given the tightness of the binary system and its symmetric light curve featuring two identical eclipsing intervals,
we assumed a circular orbit and synchronous rotation for both components.
We fixed $e$ = 0, $w$ =0, and $F_1$ = $F_2$ = 1.0 in the W-D program.
Gravity darkening coefficients ($g_1$, $g_2$) were fixed at 0.32 for the low-temperature component
with a convective atmosphere and at 1.0 for the high-temperature component with a radiative atmosphere,
following the conventions established by \citet{Luc67}.
Concerning bolometric albedos ($A_1$, $A_2$), values of 0.5 and 1.0 were typically assigned for convective and radiative atmospheres, respectively,
in accordance with principles outlined in \citet{Ruc69}.
Limb darkening coefficients were automatically interpolated from coefficient tables based on the actual surface temperature and gravity,
as detailed by \citep{van93}.
Additionally, \citet{van93} recently updated their data files, which include TESS passband (IBAND = 95) information for limb darkening coefficients.
Access to these updated data files is necessary when running the 2015 version of the W-D program\footnote{\url{https://faculty.fiu.edu/~vanhamme/lcdc2015/}}.

To determine systems' and absolute parameters simultaneously, both light and RV curves
were used as inputs in the W-D program.
During the modeling process, the following parameters  were treated as free variables:
mass ratio ($q$), orbital semi-major axis ($a$), secondary temperature ($T_2$), inclination of the system's orbit ($i$), phase shift,
surface potential $\Omega$ of either the primary or secondary component, radial velocity of the system's mass center ($V_\gamma$),
and the fractional value of the primary luminosity.

The W-D program encompasses various modes corresponding to different binary configurations.
In this study, we explored both detached and semi-detached modes for each system.
For all eight systems, both detached and semi-detached yielded reasonable fittings.
The parameters obtained from these modes are nearly identical within the specified error ranges.
The results deduced from both modes indicate that the eight systems can be classified as semi-detached binaries.
Consequently, we focus on the results from  semi-detached mode in the subsequent discussion.
 To search for the best-fitting binary model and obtain more reliable stellar parameters, the q-search method was applied to constrain the mass ratio values for spectroscopic single-lined binaries \citep{Lia12}. The step size for q was set to 0.05 - 0.1 for q values ranging from 0.05 to 1.0.
For exploring the structures of these systems better,  the fill-out factors were calculated according to the formula $f_{fill} = \Omega_I/\Omega - 1$ for $\Omega_I < \Omega$
 and $f_{fill} = (\Omega_I-\Omega)/(\Omega_I-\Omega_o)$ for $\Omega_I \geqslant \Omega$,
where $\Omega$ is the modified equipotential potential of the system, $\Omega_{I}$ is the value of the
inner critical Roche equipotential, and
$\Omega_{o}$ is the value of the outer critical Roche equipotential \citep{Bra05}.

To achieve a more realistic error estimate for $T_2$, the error
of $T_1$ was taken into account. Tests on individual light curves indicated
that $T_2$ is highly sensitive to changes in $T_1$ (also see \citet{Liakos18}).
Therefore, this errors in $T_2$ was derived from the W-D fittings based on the errors in $T_1$.
We checked the sources of contamination and incorporated their effects into the W-D program for each studied system.
Due to significant contamination in TIC 0312060302, we utilized ASAS-SN and Gaia observations instead of TESS data.
Similarly, to mitigate the impact of nearby sources on TIC 159386347, we used Kepler data.

\begin{figure*}
\begin{center}
\includegraphics[width=14cm,height=8cm,angle=0,clip]{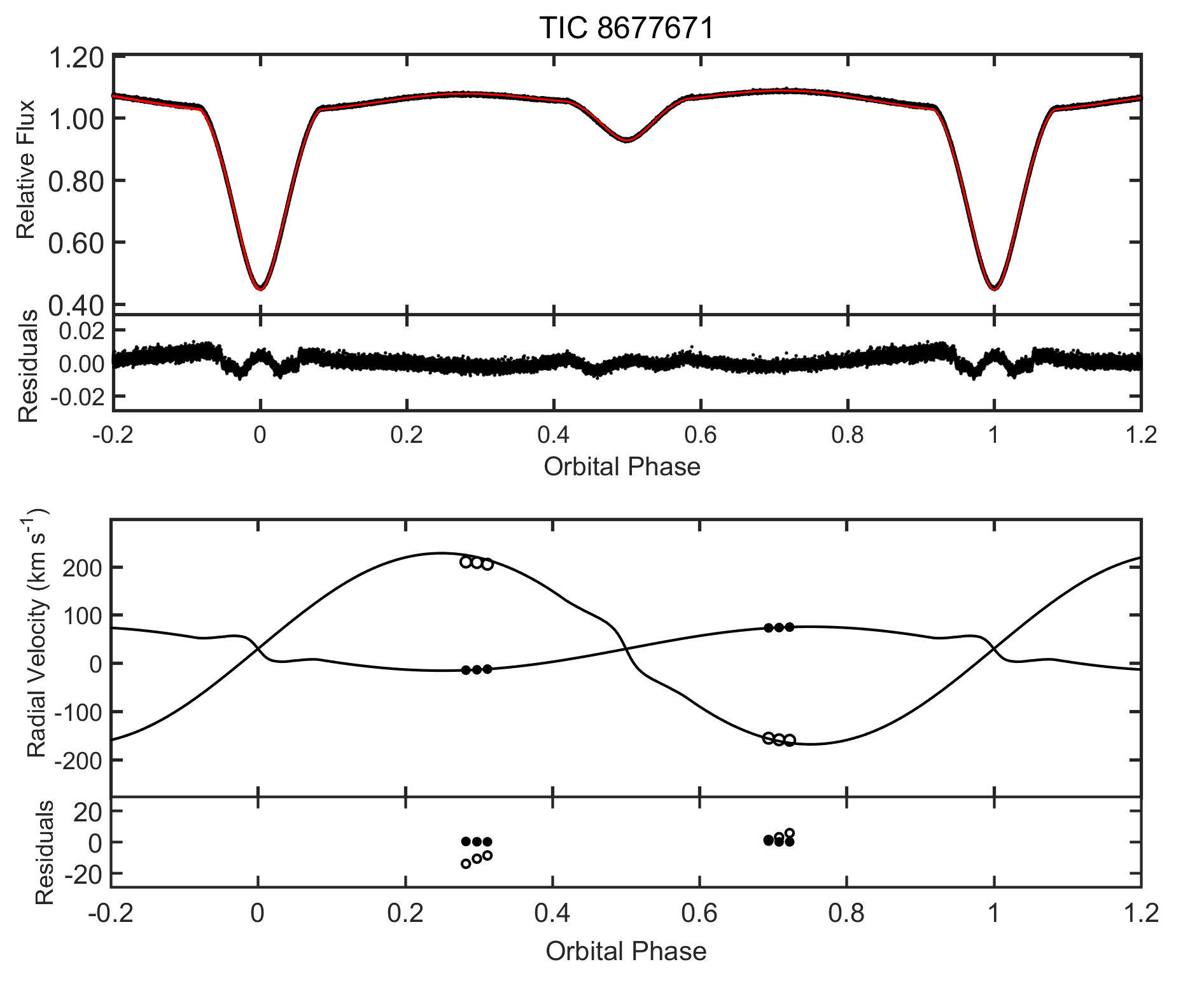}
\includegraphics[width=14cm,height=9cm,angle=0,clip]{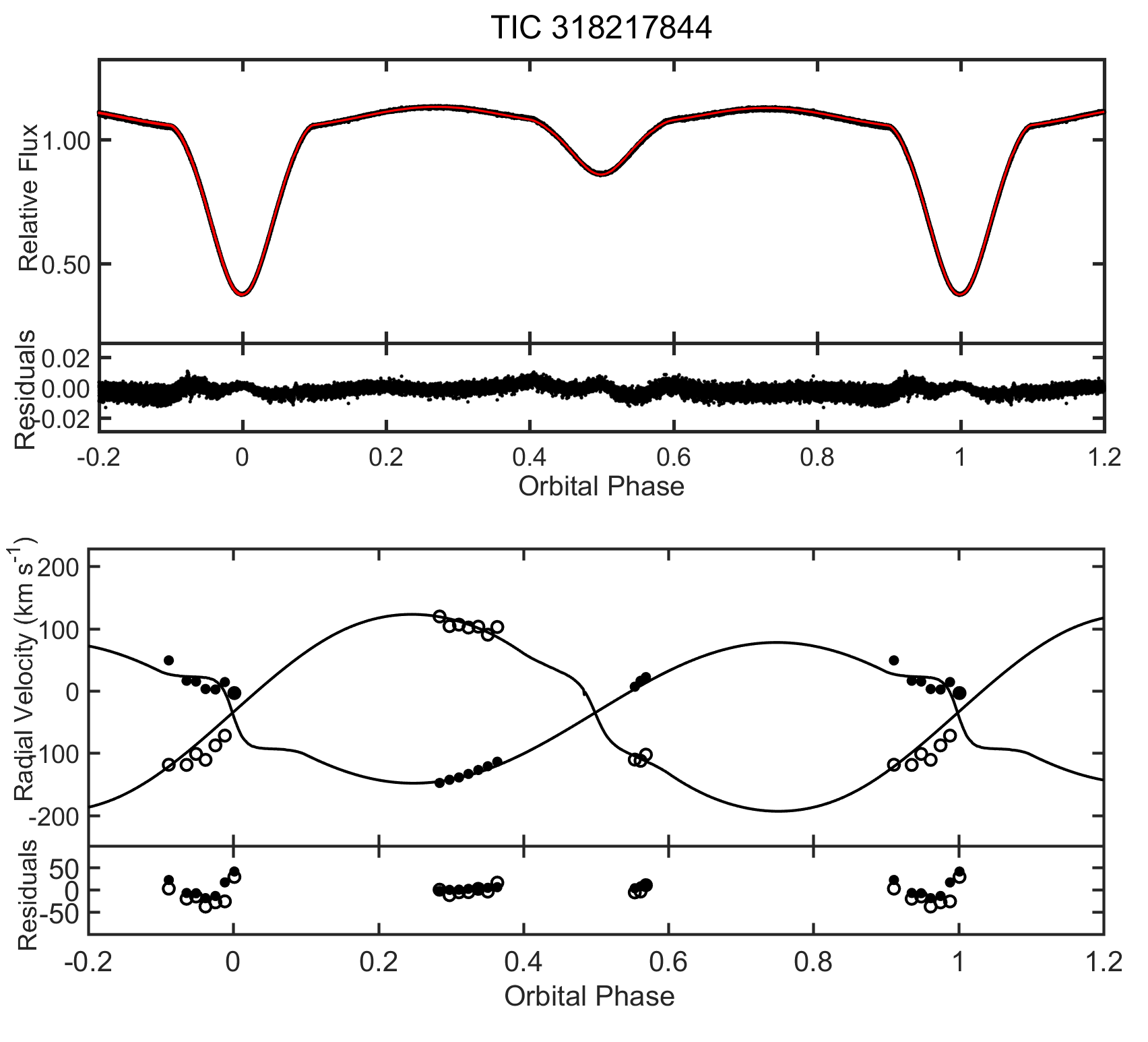}
\caption{Light and RV curves of TIC8677671 and TIC 318217844. The solid lines are theoretical fits from the W-D program.
The residuals from theoretical representation are also shown. }
\label{fig:fig3}
\end{center}
\end{figure*}

\begin{figure}
\begin{center}
\includegraphics[width=9.cm,height=7.5cm,angle=0,clip]{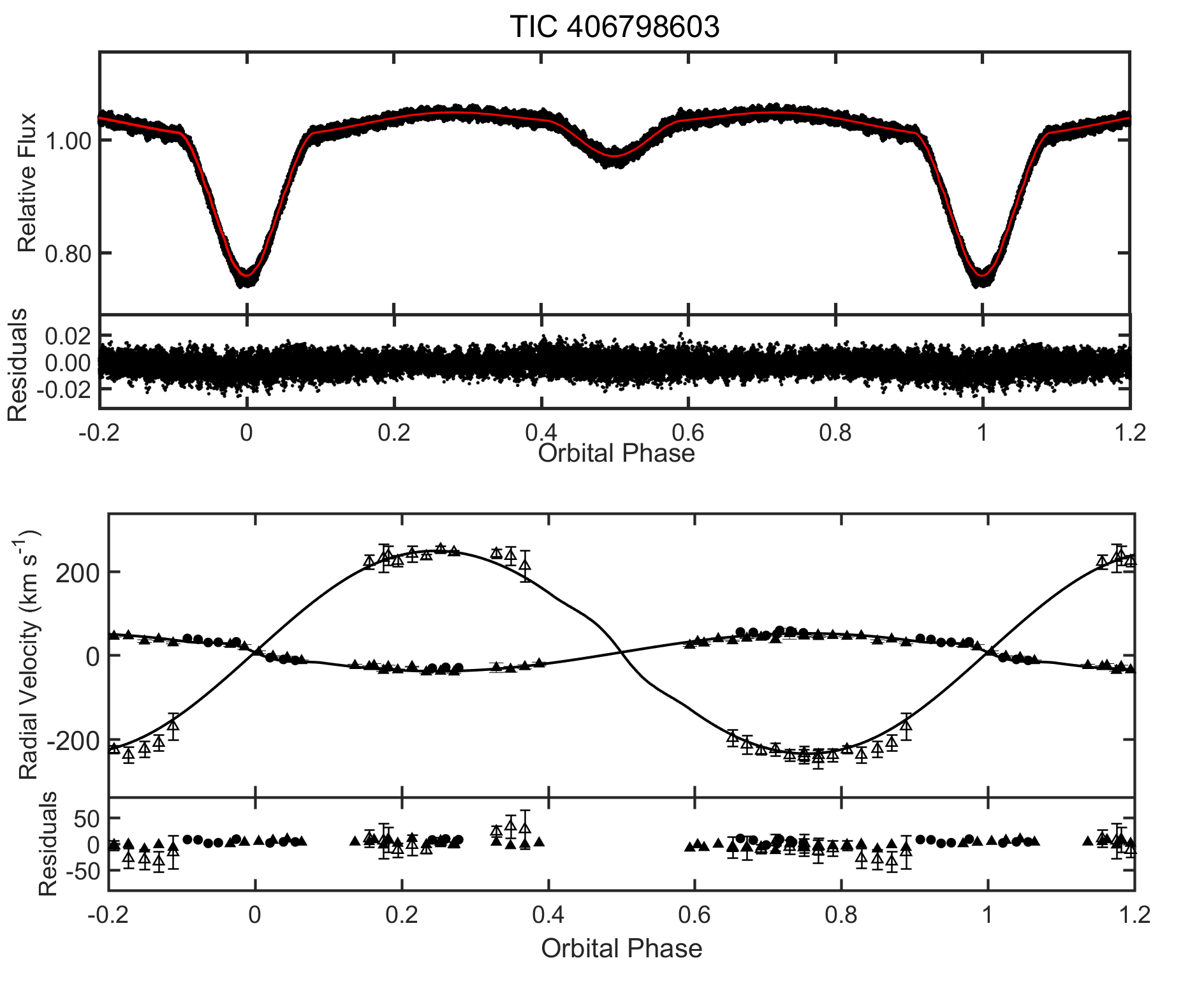}
\caption{Light and RV curves of TIC406798603. The circles and triangles represent the RV measurements from this study and from \citet{Kim22}, respectively.
The solid curves were derived from the light and RV solutions using the WD code.}
\label{fig:fig4}
\end{center}
\end{figure}

\begin{figure*}
\includegraphics[width=9.1cm,height=7.5cm,angle=0,clip]{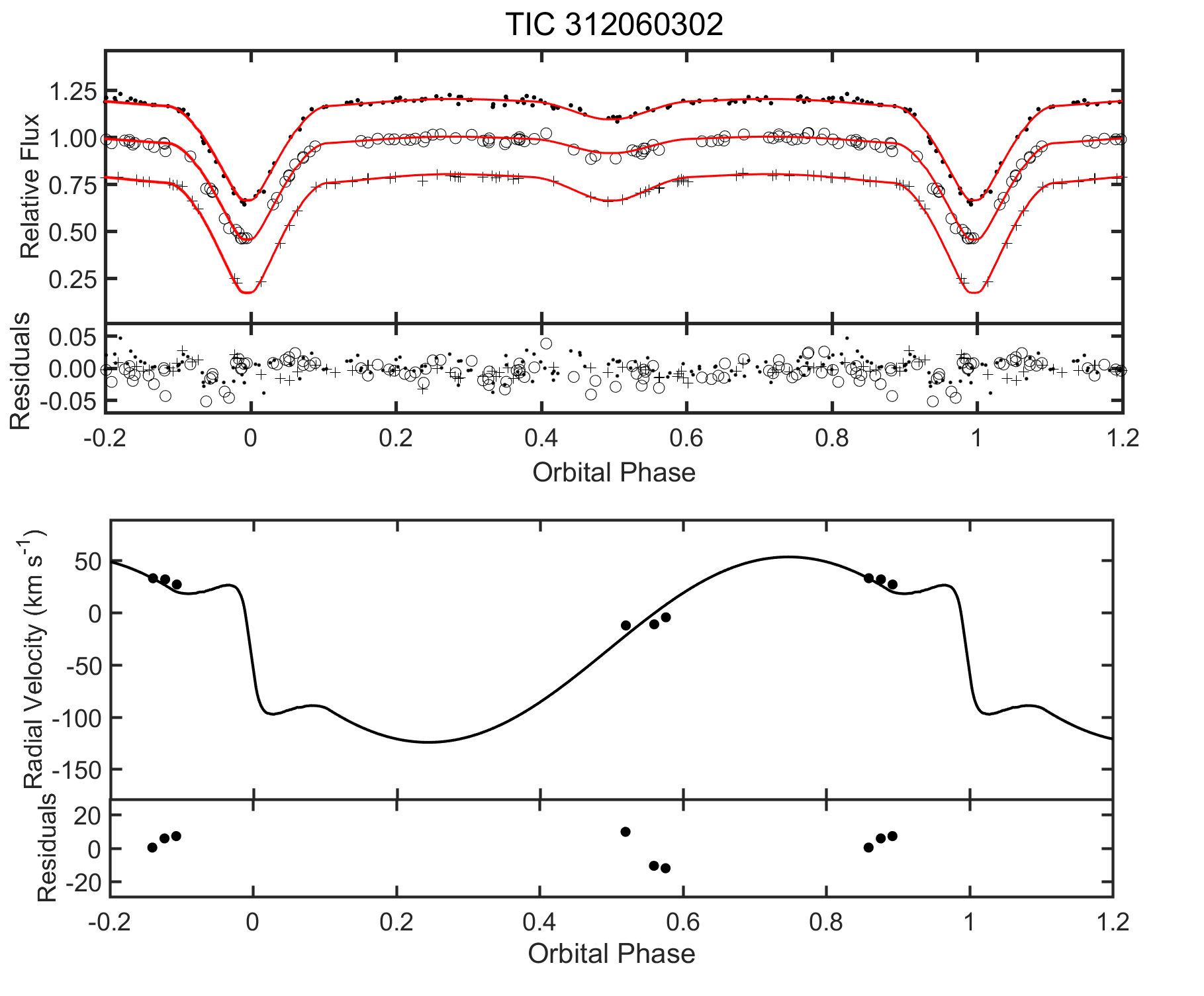}
\includegraphics[width=9.1cm,height=7.5cm,angle=0,clip]{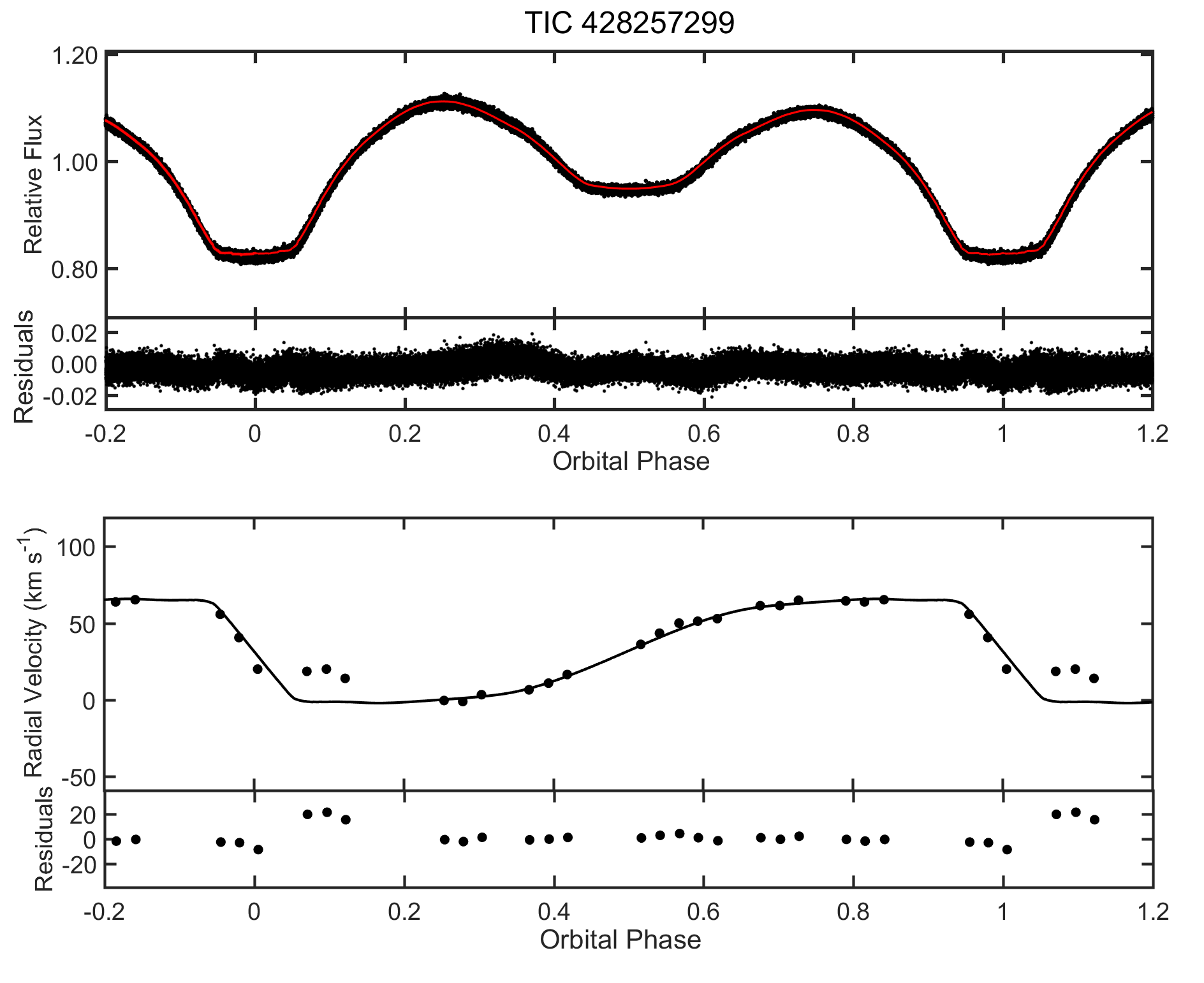}
\includegraphics[width=9.1cm,height=7.5cm,angle=0,clip]{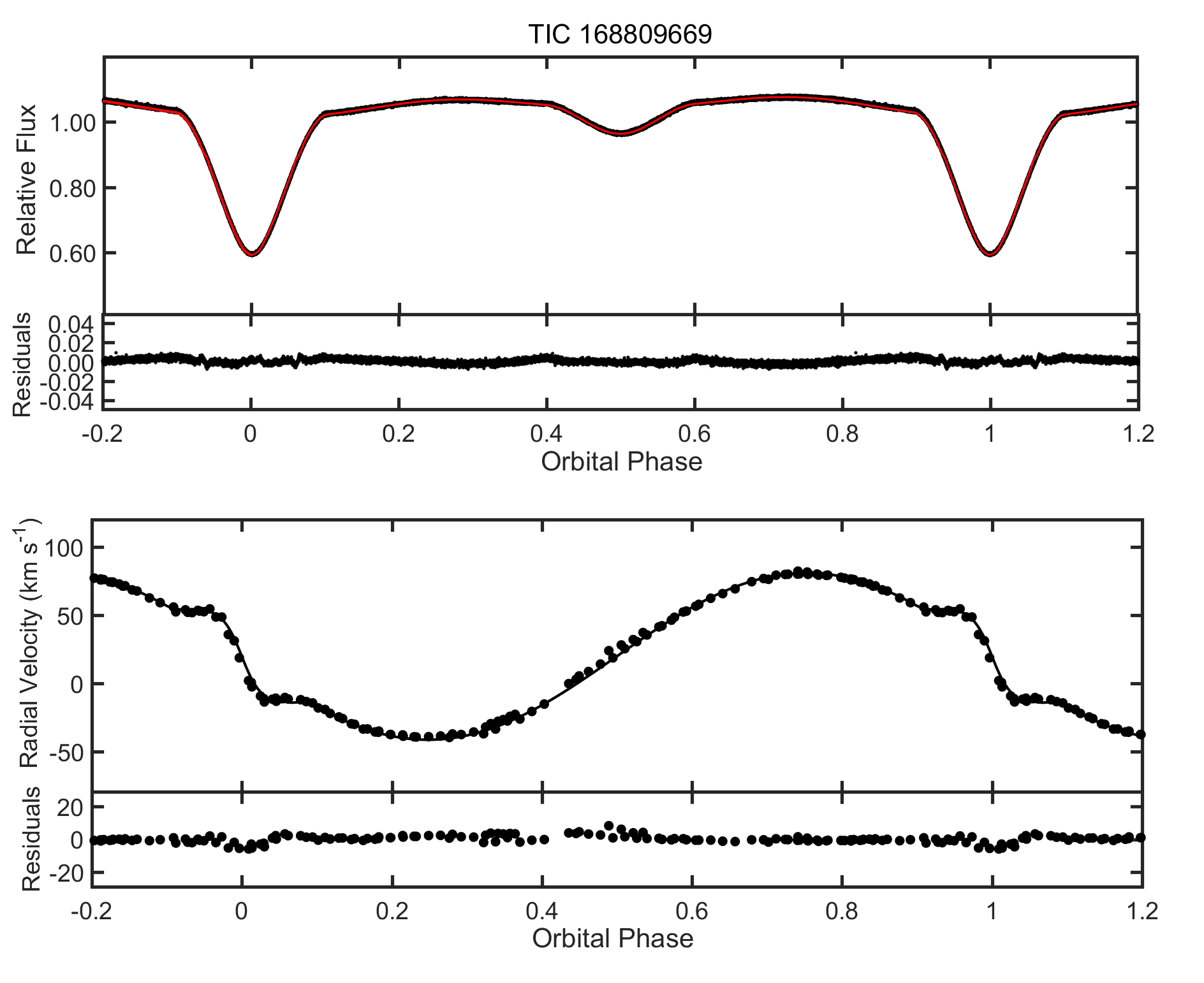}
\includegraphics[width=9.1cm,height=7.5cm,angle=0,clip]{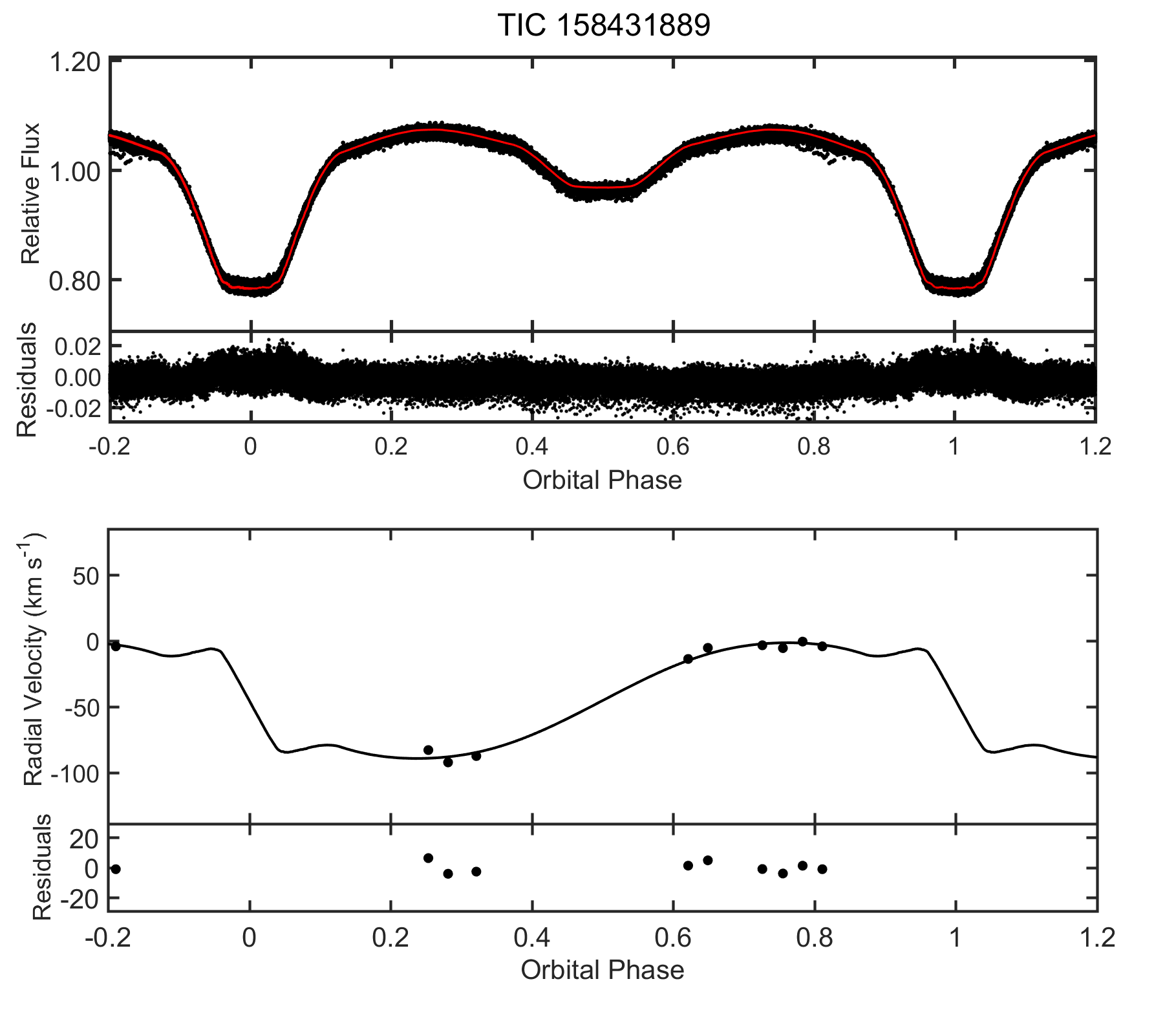}
\includegraphics[width=9.1cm,height=7.5cm,angle=0,clip]{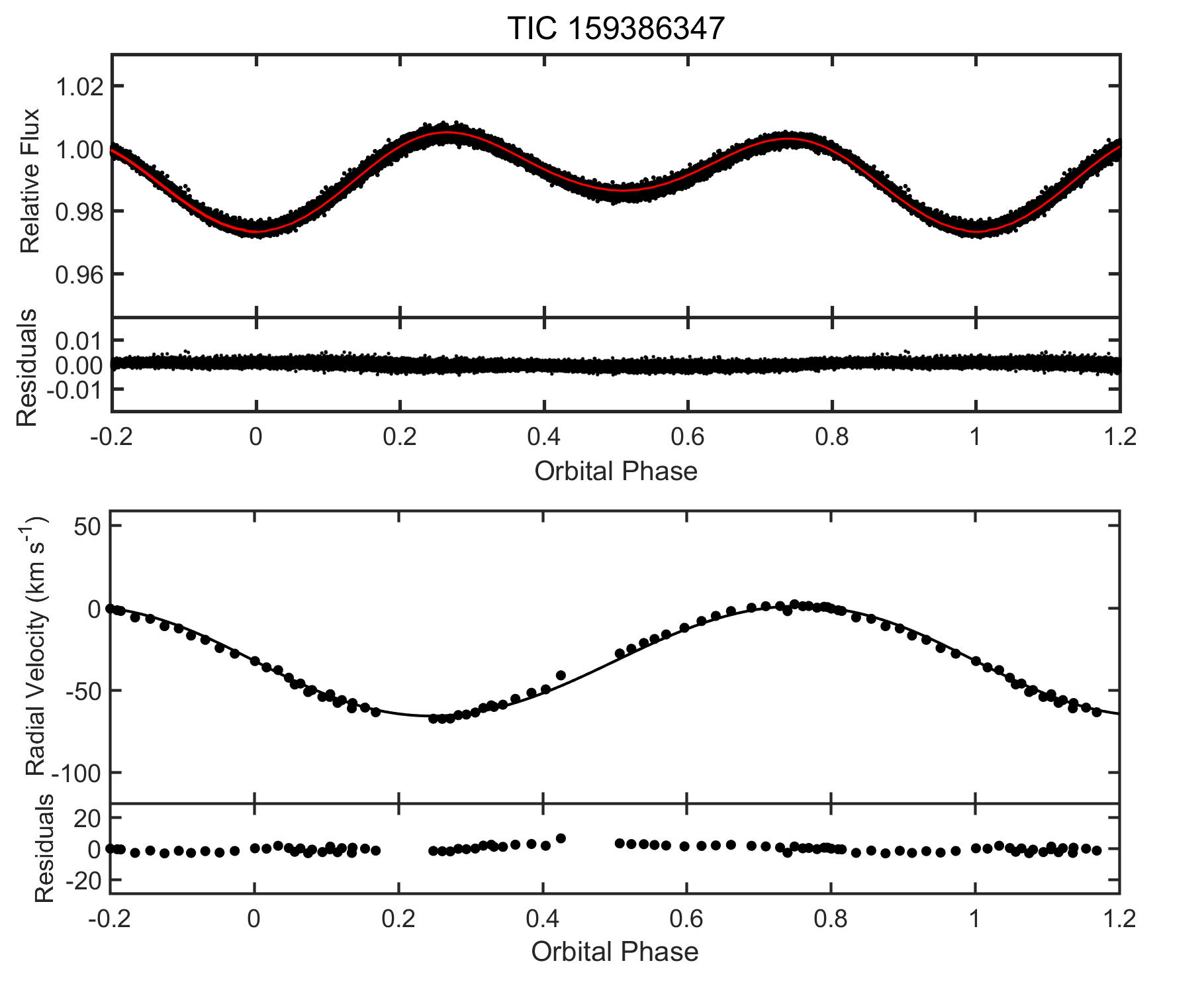}
\caption{Same as Fig. 3, just for TIC 0312060302, TIC 428257299, TIC 168809669, TIC 158431889 and TIC 159386347.}
\label{fig:fig5}
\end{figure*}

 TIC 406798603 has been identified by \citet{Kim22} as a spectroscopic double-lined binary,
we also incorporated their RV data alongside ours in the W-D fitting procedure.
The absolute parameters obtained using combined double-lined RVs and
those obtained using only our single-lined RVs are consistent within the error range.
Figure \ref{fig:fig3}, \ref{fig:fig4} and Table \ref{table:tb6}  present the geometrical and physical parameters derived from
the simultaneous analysis of the light and RV curves for  three spectroscopic double-lined binaries.
Figure \ref{fig:fig5} and Table \ref{table:tb7} correspond to the results  of  five spectroscopic single-lined binaries.
To address the asymmetries in the light curves, we included fitting models with hot spots or cold spots.
The information of these spots are given in Table \ref{table:tb8}.
The secondaries of TIC 318217844 and TIC 168809669 have best-fit albedos of 0.70 and 0.78, respectively,
which are slightly higher than the expected physical values.
The rotational distortion of the RV curve is noticeable around the phase of the primary eclipse (close to phase 0),
as depicted in TIC 318217844 of Fig.~\ref{fig:fig3}. This distortion might stem from the non-symmetric distribution of surface brightness.
The primary cause behind this non-symmetric distribution manifests as the Rossiter-McLaughlin effect.
This effect illustrates that the observed RV of a partially eclipsed star tends to exhibit a systematic bias toward either receding
or approaching velocities, depending on whether its approaching or receding hemisphere is being eclipsed \citep{Ros24, McL24}.
The results of the W-D program show that, except for TIC 428257299, where the primary star fills its Roche lobe,
the other seven systems have the secondary stars filling their Roche lobe.

\begin{table*}
\tiny
\caption{Orbital solutions for  TIC 8677671, TIC 318217844 and TIC 406798603.}
\label{table:tb6}
\centering
\begin{tabular}{l c c c}
\hline
\hline
  Parameters              &  TIC 8677671         &   TIC 318217844      &     TIC 406798603       \\
\hline
   $i^{o}$                &  83.15 $\pm$ 0.01    &  83.35 $\pm$ 0.01     &    74.04 $\pm$ 0.02      \\
  q(m$_{2}$/m$_{1}$)      &  0.221 $\pm$ 0.001   &  0.689 $\pm$ 0.001   &     0.176 $\pm$ 0.001     \\
  $T_1$(K)                &  7418 $\pm$ 160      &  9229 $\pm$ 174      &     9098 $\pm$ 174        \\
  $T_2$(K)                &  4755 $\pm$ 65       &  5762 $\pm$ 65       &     5003 $\pm$ 50        \\
  $\Omega_1$              &  4.165 $\pm$ 0.002   &  4.873 $\pm$ 0.002   &    3.111 $\pm$ 0.004      \\
  $\Omega_2$              &  2.285               &   3.224              &     2.173                \\
  $L_2/(L_1 + L_2)$       &  0.15                & 0.24                 &      0.04                \\
  $r_1$(pole)             & 0.2531 $\pm$ 0.0002  & 0.2368 $\pm$ 0.0001  &    0.3397 $\pm$ 0.0004    \\
  $r_1$(point)            & 0.2575 $\pm$ 0.0002  & 0.2439 $\pm$ 0.0001  &    0.3538 $\pm$ 0.0004   \\
  $r_1$(side)             & 0.2557 $\pm$ 0.0002  & 0.2395 $\pm$ 0.0001  &    0.3481 $\pm$ 0.0004    \\
  $r_1$(back)             & 0.2570 $\pm$ 0.0002  & 0.2426 $\pm$ 0.0001  &    0.3514 $\pm$ 0.0004    \\
  $r_2$(pole)             & 0.2397 $\pm$ 0.0001  & 0.3258 $\pm$ 0.0001  &    0.2246 $\pm$ 0.0003    \\
  $r_2$(point)            & ...                  & ...                  &    ...       \\
  $r_2$(side)             & 0.2494 $\pm$ 0.0001  & 0.3409 $\pm$ 0.0001  &   0.2335 $\pm$ 0.0003      \\
  $r_2$(back)             & 0.2819 $\pm$ 0.0001  & 0.3728 $\pm$ 0.0001  &   0.2657 $\pm$ 0.0003    \\
  $K_1$ (km $s^{-1}$)     & 45.6 $\pm$ 2.8       & 113.4 $\pm$ 5.7      &    43.6 $\pm$ 2.0        \\
  $K_2$ (km $s^{-1}$)     & 190.8 $\pm$ 3.1      & 167.5 $\pm$ 6.8      &   255.2 $\pm$ 5.5        \\
  $\gamma$ (km $s^{-1}$)  & 29.9  $\pm$ 2.6      & -35.9 $\pm$ 3.2      &   6.9  $\pm$ 1.7         \\
  $f_{fill} 1$            &  -0.52               & -0.34               &      -0.30      \\
  $f_{fill} 2$            &  0                   & 0                   &    0       \\
 \hline
 \multicolumn{4}{c}{Absolute parameters }\\
 \hline
  a(R$_{\odot}$)          & 5.59 $\pm$ 0.29      & 6.76 $\pm$ 0.24   &   5.42 $\pm$ 0.21         \\
  $M_1$(M$_{\odot}$)      & 1.51 $\pm$ 0.23      & 1.64 $\pm$ 0.18   &   2.39 $\pm$ 0.28          \\
  $M_2$(M$_{\odot}$)      & 0.33 $\pm$ 0.06      & 1.13 $\pm$ 0.17   &   0.42 $\pm$ 0.06         \\
  $R_1$(R$_{\odot}$)      & 1.42 $\pm$ 0.07      & 1.63 $\pm$ 0.06   &   1.93 $\pm$ 0.08         \\
  $R_2$(R$_{\odot}$)      & 1.43 $\pm$ 0.07      & 2.35 $\pm$ 0.09   &   1.35 $\pm$ 0.05          \\
  Log $g_1$(cgs)          & 4.31 $\pm$ 0.09      & 4.23 $\pm$ 0.07   &   4.25 $\pm$ 0.07          \\
  Log $g_2$(cgs)          & 3.65 $\pm$ 0.09      & 3.75 $\pm$ 0.07   &   3.80 $\pm$ 0.07           \\
  $L_1$(L$_{\odot}$)      & 5.48 $\pm$ 0.74      & 17.23 $\pm$ 1.80  &   22.90 $\pm$ 2.52          \\
  $L_2$(L$_{\odot}$)      & 0.94 $\pm$ 0.11      & 5.45 $\pm$ 0.46   &    1.02 $\pm$ 0.09          \\
\hline
\end{tabular}
\end{table*}

\begin{table*}
\tiny
\caption{Orbital solutions for five spectroscopic single-lined binaries.}
\label{table:tb7}
\centering
\begin{tabular}{l c c c c c }
\hline
\hline
  Parameters           & TIC 0312060302     & TIC 428257299        &  TIC 168809669         &   TIC 158431889             &  TIC 159386347         \\
\hline
   $i^{o}$             & 86.97 $\pm$ 0.53   & 87.01 $\pm$ 0.02     &  78.72 $\pm$ 0.03      &  86.36 $\pm$ 0.02           &  52.10 $\pm$ 0.07       \\
  q(m$_{2}$/m$_{1}$)   & 0.304 $\pm$ 0.007  & 0.163 $\pm$ 0.001    &  0.272 $\pm$ 0.001     &  0.145 $\pm$ 0.001          &  0.160 $\pm$ 0.001       \\
  $T_1$(K)             & 11481 $\pm$ 293    & 7104 $\pm$ 168       &  7988 $\pm$ 181        &  7718 $\pm$ 117             &  7628 $\pm$ 126          \\
  $T_2$(K)             & 6076  $\pm$ 80     & 4676 $\pm$ 101       &  4294 $\pm$ 60         &  4689 $\pm$ 47              &  3946 $\pm$ 116        \\
  $\Omega_1$           & 3.287 $\pm$ 0.018  & 2.122 $\pm$ 0.001    &  3.272 $\pm$ 0.004     &  2.367 $\pm$ 0.001          &  2.771 $\pm$ 0.001      \\
  $\Omega_2$           &  2.475             & 2.152 $\pm$ 0.001    &  2.404                 &  2.091                      &  2.129                   \\
  $L_2/(L_1 + L_2)$    & 0.05               & 0.03                 &   0.05                 & 0.03                        & 0.03                     \\
  $r_1$(pole)          &0.3330 $\pm$ 0.0022 &  0.5058 $\pm$ 0.0001 & 0.3315 $\pm$ 0.0004    & 0.4476 $\pm$ 0.0001         &  0.3813 $\pm$ 0.0004    \\
  $r_1$(point)         &0.3509 $\pm$ 0.0029 &  ...                 & 0.3478 $\pm$ 0.0005    & 0.4981 $\pm$ 0.0002         &  0.4043 $\pm$ 0.0005    \\
  $r_1$(side)          &0.3416 $\pm$ 0.0025 &  0.5554 $\pm$ 0.0002 & 0.3397 $\pm$ 0.0004    & 0.4747 $\pm$ 0.0002         &  0.3948 $\pm$ 0.0004    \\
  $r_1$(back)          &0.3470 $\pm$ 0.0027 &  0.5777 $\pm$ 0.0002 & 0.3445 $\pm$ 0.0005    & 0.4844 $\pm$ 0.0002         &  0.3998 $\pm$ 0.0004    \\
  $r_2$(pole)          &0.2620 $\pm$ 0.0015 &  0.2154 $\pm$ 0.0005 & 0.2545 $\pm$ 0.0003    & 0.2123 $\pm$ 0.0001         &  0.2181 $\pm$ 0.0002    \\
  $r_2$(point)         &  ...               &  0.2834 $\pm$ 0.0026 & ...                    & ...                         &  ...                   \\
  $r_2$(side)          &0.2729 $\pm$ 0.0016 &  0.2235 $\pm$ 0.0006 & 0.2649 $\pm$ 0.0003    & 0.2207 $\pm$ 0.0001         &  0.2268 $\pm$ 0.0002    \\
  $r_2$(back)          &0.3056 $\pm$ 0.0016 &  0.2522 $\pm$ 0.0010 & 0.2975 $\pm$ 0.0003    & 0.2525 $\pm$ 0.0001         &  0.2588 $\pm$ 0.0002    \\
  $K_1$ (km $s^{-1}$)   & 90.5 $\pm$ 16.6   &  35.3 $\pm$ 1.5      & 61.9 $\pm$ 1.0         & 43.9 $\pm$ 3.1              &  33.7 $\pm$ 0.5         \\
  $\gamma$ (km $s^{-1}$)& -35.6 $\pm$ 11.9  &  31.9 $\pm$ 0.8      & 23.9 $\pm$ 0.7         & -45.1 $\pm$ 1.3             &  -32.5 $\pm$ 0.3       \\
  $f_{fill} 1$          &  -0.26            & 0.15                 &  -0.27                 &   -0.12                     &  -0.23                 \\
  $f_{fill} 2$          &  0                & -0.001               &      0                 &   0                         &   0                   \\
 \hline
 \multicolumn{6}{c}{Absolute parameters }\\
 \hline
  a(R$_{\odot}$)       &  7.53 $\pm$ 1.11   & 3.18 $\pm$ 0.12      & 5.86 $\pm$ 0.08        & 3.93 $\pm$ 0.24             &  4.94 $\pm$ 0.06         \\
  $M_1$(M$_{\odot}$)   &  4.40 $\pm$ 1.94   & 0.91 $\pm$ 0.10      & 2.10 $\pm$ 0.08        & 2.33 $\pm$ 0.44             &  2.16 $\pm$ 0.08       \\
  $M_2$(M$_{\odot}$)   &  1.34 $\pm$ 0.77   & 0.15 $\pm$ 0.02      & 0.57 $\pm$ 0.03        & 0.34 $\pm$ 0.07             &  0.35 $\pm$ 0.02        \\
  $R_1$(R$_{\odot}$)   &  2.54 $\pm$ 0.37   & 1.74 $\pm$ 0.06      & 1.99 $\pm$ 0.03        & 1.89 $\pm$ 0.12             &  1.94 $\pm$ 0.03       \\
  $R_2$(R$_{\odot}$)   &  2.09 $\pm$ 0.31   & 0.73 $\pm$ 0.03      & 1.60 $\pm$ 0.02        & 0.92 $\pm$ 0.06             &  1.17 $\pm$ 0.02       \\
  Log $g_1$(cgs)       &  4.27 $\pm$ 0.28   & 3.92 $\pm$ 0.06      & 4.16 $\pm$ 0.03        & 4.25 $\pm$ 0.10             &  4.19 $\pm$ 0.02         \\
  Log $g_2$(cgs)       &  3.92 $\pm$ 0.28   & 3.88 $\pm$ 0.06      & 3.79 $\pm$ 0.03        & 4.04 $\pm$ 0.10             &  3.84 $\pm$ 0.02         \\
  $L_1$(L$_{\odot}$)   &  100.6 $\pm$ 31.3  & 6.89 $\pm$ 0.83      & 14.42 $\pm$ 1.36       & 11.37 $\pm$ 1.58            &  11.43 $\pm$ 0.81        \\
  $L_2$(L$_{\odot}$)   &  5.34 $\pm$ 1.60   & 0.23 $\pm$ 0.03      & 0.78 $\pm$ 0.05        & 0.37 $\pm$ 0.05             &  0.30 $\pm$ 0.04        \\
\hline
\end{tabular}
\end{table*}

\begin{table*}
\caption{Information of spots  in the best-fitting model.
The errors  are determined through visual inspection of their effects on the light curve.}
\label{table:tb8}
\centering
\begin{tabular}{l c c c c}
\hline
\hline
                                &  TIC 8677671   &  TIC 318217844  &   TIC 428257299  &  TIC 168809669         \\
\hline
Component                       &   Primary      & Primary         & Secondary        &     Primary             \\
Co-latitude                     & $90.1^\circ$ $\pm$ 2.0   & $154.8^\circ$ $\pm$ 2.0    & $67.2^\circ$ $\pm$ 1.0      &   $80.2^\circ$ $\pm$ 2.0           \\
Longitude                       & $158.9^\circ$ $\pm$ 2.0  & $130.8^\circ$ $\pm$ 2.0    & $257.8^\circ$ $\pm$ 5.0     &   $200.5^\circ$ $\pm$ 4.0           \\
Radius                          & $56.7^\circ$ $\pm$ 1.0    & $40.1^\circ$ $\pm$ 1.0     & $43.0^\circ$ $\pm$ 1.0      &   $60.7^\circ$ $\pm$ 1.0           \\
$T_{\rm spot}$ / $T_{\rm star}$ &       1.014 $\pm$ 0.001    & 0.97 $\pm$ 0.01            & 0.85 $\pm$ 0.02             &   1.06 $\pm$ 0.01                  \\

\hline
\end{tabular}
\end{table*}

\section{Discussions}

\subsection{Evolutionary states}

We conducted an analysis on eight semi-detached eclipsing binaries,
determining their absolute parameters through simultaneous fitting of high-precision photometric data primarily
from TESS and RV curves obtained from LAMOST MRS.
Among these systems,  three are spectroscopic  double-lined binary systems, and  five are spectroscopic  single-lined binary systems.
In the case of TIC 428257299, the primary star fills its Roche lobe.
However, in the remaining seven systems, their secondaries fill their Roche lobes, while the primaries remain detached from the lobes.
Notably, among these eight systems, the absolute parameters of TIC 0312060302, TIC 428257299, TIC 158431889, and TIC 159386347
are calculated for the first time in this work, and the absolute parameters of TIC 8677671 and TIC 168809669 are determined
for the first time using combined photometric and spectroscopic data in this study.
TIC 318217844 and TIC 406798603 have been previously analyzed in other works, combining photometric and spectroscopic data \citep{Dou15, Kim22, Lee24}.

The system's parameters of TIC 158431889 and TIC 159386347 had been estimated by \citet{Prs11, Sla11} and \citet{Arm14}, which based solely on photometric data.
They reported significantly different parameters for both TIC 158431889 and TIC 159386347
but arrived at the same conclusion of morphology types.
The absolute parameters of TIC 8677671, TIC 168809669 and TIC 318217844 were calculated  relied on older or lower quality data \citep{Bra80, Bud04, Sve90}.
\citet{Man15} calculated the absolute parameters of TIC 168809669 solely using photometric data.
In comparison to their findings, our parameters are derived from simultaneous modeling of light and RV curves using the binary W-D program.
For TIC 318217844, \citet{Lee24} presented the absolute properties, utilizing data from TESS and Bohyunsan Observatory Echelle Spectrograph.
Within the margin of error, our results are close to theirs.


\citet{Dou15} and \citet{Kim22} provided the absolute parameters of TIC 406798603 through
an analysis of photometric and high-resolution spectroscopic data.
Additionally, \citet{Kim22} delved into further discussions on secular variations of orbital period.
Since distinct double lines of  TIC 406798603 were not resolved in the spectra obtained from LAMOST,
 we  list the results calculated  by combining  RV data from \citet{Kim22} and ours.

\begin{figure*}
\begin{center}
\includegraphics[width=13.8cm,height=7cm,angle=0,clip]{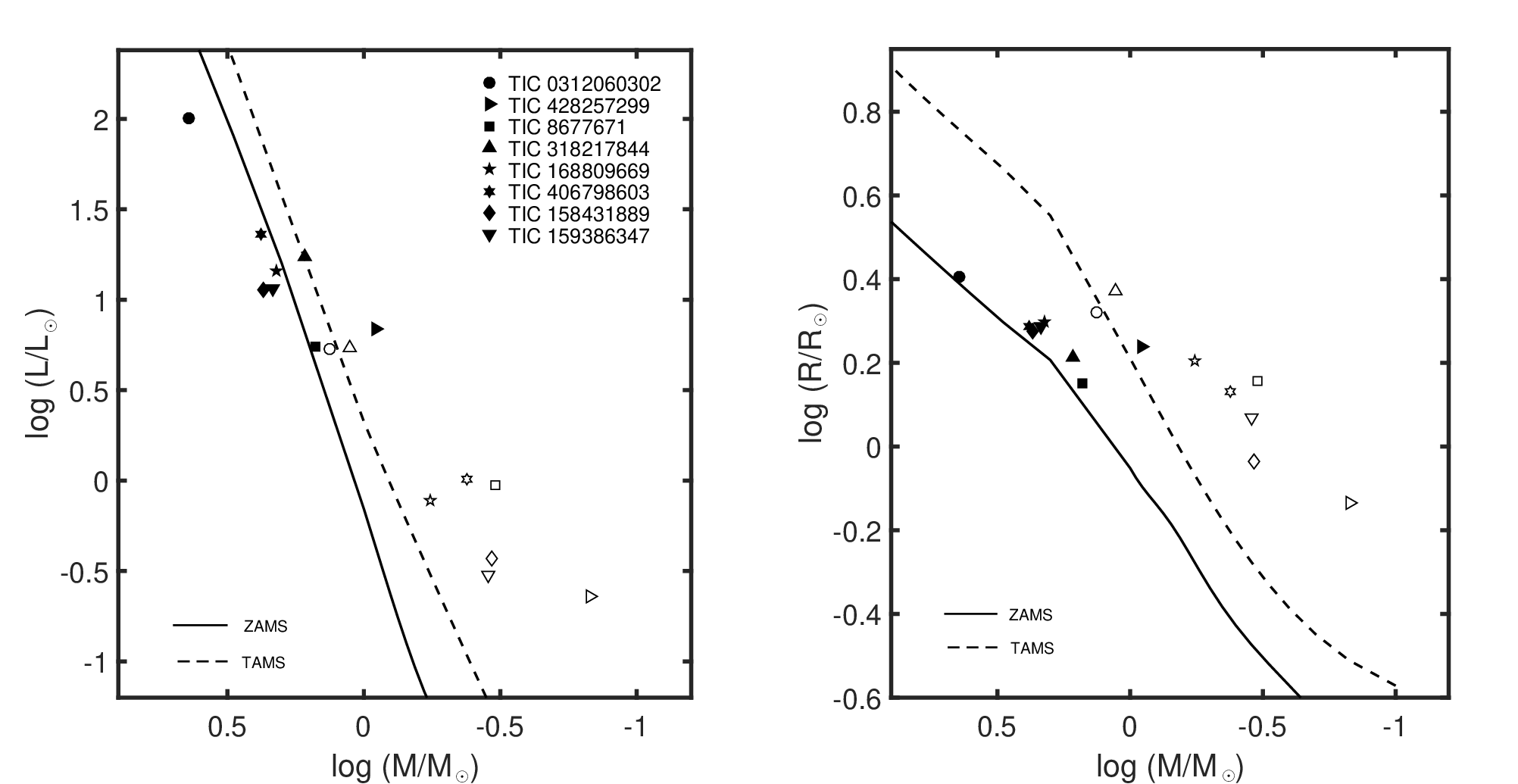}
\caption{M-L and M-R relations of the eight systems. The primaries and the secondaries are represented by filled and open symbols, respectively.
The ZAMS and the TAMS that obtained using the Binary Star Evolution code from \citet{Hur02} are described with solid and dashed black lines. }
\label{fig:fig6}
\end{center}
\end{figure*}

In order to analyze the evolutionary states of the eight systems,
the absolute parameters of both components of these eclipsing binaries are presented in logM-logL and logM-logR diagrams (Fig.~\ref{fig:fig6}).
The zero-age main sequence (ZAMS) and the terminal-age main sequence (TAMS)
were derived using the Binary Star Evolution code of \citet{Hur02}.
Except for TIC 428257299, the primary stars in the other seven systems are located within the main-sequence band,
indicating that they are  non-evolved  main-sequence stars.
Conversely, the secondaries have evolved near or beyond the TAMS, appearing oversized and overluminous compared to main-sequence stars of the same mass. They are classified as subgiants with lower mass than the primaries.
These results suggest that these binaries might have formed from initially detached binary systems
through case A mass transfers, subsequently undergoing mass-reverse evolutions \citep{Pol94, Qia13}.
Additionally, these binaries are probably still in the slow mass exchange stage,
with mass transferring from the less massive secondaries to the more massive primaries.

Regarding TIC 428257299, the primary star fills its Roche lobe.
The secondary star is approaching Roche lobe overflow
and has a mass lower than that of a main-sequence star with the same temperature,
implying previous mass loss. It might have undergone a phase of thermal relaxation oscillation,
recently experienced contact without reaching the thermal contact phase, and subsequently been disrupted into its current semi-detached state.
At present, the secondary star exhibits a larger radius compared to a main-sequence star of equivalent mass,
indicating the possibility of previous experiences with mass ratio reversal and contact processes.
Figure~\ref{fig:fig6} illustrates the evolution of both primary and secondary components.

In addition, among these systems, TIC 318217844 exhibits a high mass ratio of  $m_2/m_1 = 0.689$
and a radius ratio of  $R_2/R_1\sim 1.44$.
This finding aligns with the observations of \citet{Li22}, which demonstrated a correlation between mass ratios and radii ratios for semi-detached binaries in M31.
In their sample, comprising 12 semi-detached systems,
only two exhibited high mass ratios (0.596 and 0.843) and radius ratios ($R_2/R_1$ $> 1$).
This suggests the necessity for further study involving larger samples to comprehensively understand the relationship
between mass ratios and radii ratios of semi-detached binaries.

\subsection{Long-term period variations}

To investigate the evolutionary phases of these binaries, we analyzed their orbital period variations and identified secular changes in the orbital periods of TIC 8677671, TIC 318217844,
and TIC 406798603. As TIC 406798603 has already been studied in the work of \citet{Kim22}, we do not discuss it here.
 The cyclic oscillations of the orbital periods are observed in TIC 8677671 and TIC 318217844; however, a quadratic term is evident only in TIC 318217844 (Fig. 1).
Mass transfer from the secondary star to the primary star results in an increase in orbital period.
Therefore, the observed decrease in orbital period in TIC 318217844 may be attributed to the loss of mass and angular momentum.

For the late-type binaries, the cyclic oscillations of the orbital periods are generally explained by two mechanisms: magnetic activity cycles \citep{App92} or light-time effects due to the presence of a third body \citep{Irw52}.
\citet{App92} suggested that the observed amplitude of period modulation could come from the gravitational coupling mechanism caused by the cyclic magnetic activity.
Therefore, the value of $\Delta$P/P was computed according to the following equation \citep[also see][]{Yan21},

\vspace*{-0.1 cm}
\begin{equation}
{\triangle P \over P}  = {A \times {\sqrt{2[1-cos({{2\pi P} \over {P_{3}}})]}}} \simeq {{2\pi A} \over {P_{3}}} ,
\end{equation}

where A is $a_{12}sin$i$/c$. For TIC 8677671, the two values of  $\triangle P/P$ are  approximately  $\sim$ 2 $\times$ $10^{-5}$  and $\sim$ 1 $\times$ $10^{-5}$,
consistent to the typical value of the magnetic activity mechanism.
In TIC 318217844, $\triangle P/P$ $\sim$ 3 $\times$  $10^{-6}$, is slightly below the typical value of the magnetic activity mechanism.
The magnetic activity mechanism is considered a plausible cause for the cyclic changes in orbital periods observed in both systems.

The another mechanism which causes the cyclic change of the orbital periods is
the light-time effect via by the presence of the third body.
Its mass function can be computed by the following equation:

\vspace*{-0.1 cm}
\begin{equation}
f(m) = {{{4\pi^2}\over{G P_3^2}} \times {(a_{12}sini)^3}} = {{M_3sini}^3 \over {(M_1+M_2+M_3)^2}} .
\end{equation}

We computed the parameters about the third body by using the program provided by \citet{Zas09}.
the data is weighted according to their observing techniques,
namely visual observations  are weighted with 1, photographic observations are weighted with 5 and CCD \& Photoelectric
observations are weighted with 10.
The results show that the minimum mass $M_3$ of TIC 318217844 is  about 0.42 $M_\odot$  for the orbital inclination $i$ = $90^\circ$.
Assuming that the third body is a main-sequence star, its luminosity is about 0.04 $L_\odot$.
Therefore, it is difficult to identify the third body due to the low luminosity.
Since the third light was not found in orbital solutions,
we tend to think the O-C change is caused by magnetic activity.

For TIC 8677671, cyclic changes are modeled with two eccentric orbits.
One has an orbital period of about 97.9 yrs and projected semi-major axis of 22.3 AU indicating a mass function of $f(m_3)$ = 1.15 corresponding to
a mass of about 2.97 M$_{\odot}$ third body for the orbital inclination $i$ = $90^\circ$ in the system.
The other an orbital period of about 15.89 yrs and projected semi-major axis of 1.54 AU indicating a mass function of $f(m_4)$ = 0.0145,
corresponding to  a mass of about 0.41 M$_{\odot}$  body for the orbital inclination $i$ = $90^\circ$ in the system.
The primary of TIC 8677671 is an F-type main-sequence star.
The minimum mass of the third body is greater than the total mass of the TIC 8677671 system,
i.e., $M_3$ $>$ $M_1$ + $M_2$.
If this third body lies on the main sequence, its luminosity should exceed that of the primary star in TIC 8677671, making it observable. However, spectral observations do not detect this object \citep{Qian08}.
When we treated the third light as a free parameter in the W-D code, the contribution of the third light was found to be about 6\% of the total luminosity. Approximately 9 arcseconds away from TIC 8677671, there is a source (Gaia DR3 699818436453891072) with a brightness approximately 4 magnitudes fainter than that of TIC 8677671.
TIC 8677671 is located at a distance of 276 pc, while the fainter source is located at a distance of 262 pc.
Contaminated light from this source (Gaia DR3 699818436453891072) contributes roughly 3\% to the overall luminosity (see Table 3).
This suggests that the third body is invisible at optical wavelengths, implying that the unseen third body could be a stellar-mass black hole candidate.
Additionally, a faint X-ray source (1RXS J090950.1+302514) located near TIC 8677671 was detected by ROSAT.
It has a count rate of 0.016 $\pm$ 0.008 counts s$^{-1}$ in the 0.1-2.4 keV energy band \citep{Voges00}.
The third companion may correspond to this weak X-ray source.

\section{Conclusions}

Semi-detached binaries are important to understand the binary evolution because
they are the intermediate phase from detached to the contact phase for close interacting binary stars.
In this work, we have analyzed 8 semi-detached  binaries by simultaneously modeling the light curves
and RV curves with the help of the W$-$D program. The absolute parameters in each binary have been well derived.
Their structures and  evolutionary states  are also studied.
We also analyzed the secular variation of orbital periods of two systems.
The results are summarized as follows:

\begin{enumerate}
\item
All the eight systems are semi-detached binaries with mass ratios from  0.145 to 0.689.
For TIC 0312060302, TIC 428257299, TIC 158431889 and TIC 159386347,
their absolute parameters are provided in this work for the first time.
For TIC 8677671 and TIC 168809669, their absolute parameters are calculated by simultaneously modeling the light curves
and RV curves for the first time.
Except of TIC 428257299 which the primary fills the Roche lobe,
the  secondaries of the other seven systems fill their Roche lobes.

\item We have analyzed the evolutionary states of both components of each system based on their absolute parameters.
Except for TIC 428257299, the primary stars of the other systems are located in the main-sequence band, but the secondaries are all evolved away from the main-sequence band.
This result indicates that they probably formed from originally detached binary systems and experienced the mass-reverse evolution.
The primary of TIC 428257299 is filling its Roche lobe, and the companion star had undergone a process of mass loss.

\item The orbital periods of \emph{TIC 318217844} and \emph{TIC 8677671} show secular cyclic oscillations.
The secular periodic changes in TIC 318217844 may be caused by magnetic activity.
However, for TIC 8677671, the secular periodic changes may be caused by  magnetic activity or a third body, which appears to be a compact object.

\end{enumerate}

\begin{acknowledgements}

The authors thank the anonymous referee for the very helpful suggestions.
The work is supported by the National Key R\&D Program of China(2019YFA0405000),
the Strategic Priority Research Program of the Chinese Academy of Sciences (grant No. XDB41000000),
Key Research Program of Frontier Sciences, CAS (grant No. ZDBS-LY-SLH013),
Frontier Scientific Research Program of Deep Space Exploration Laboratory (2022-QYKYJH-ZYTS-016),
and the National Natural Science Foundation of China (grant Nos. 12373111, 12350004 and 12273018).
\par
Guoshoujing Telescope (the Large Sky Area Multi-Object Fiber Spectroscopic Telescope LAMOST) is a National Major Scientific Project built
by the Chinese Academy of Sciences. Funding for the project has been provided by the National Development and Reform Commission.
LAMOST is operated and managed by the National Astronomical Observatories, Chinese Academy of Sciences.
\par
\end{acknowledgements}

%
\bibliographystyle{aa} 
\bibliography{ref}
\newpage

\appendix

\section{Radial velocities of the eight binaries.}

\begin{table}[ht]
\caption{Radial velocities of TIC 8677671 and TIC 318217844.}
\label{table:A1}
\centering
\begin{tabular}{l c c c c }
\hline\hline
BJD-2457000        & $RV_{1}$ (km $s^{-1}$)  & $\Delta RV_{1}$ (km $s^{-1}$) &  $RV_{2}$ (km $s^{-1}$) & $\Delta RV_{2}$ (km $s^{-1}$)\\
\hline
\multicolumn{5}{c}{TIC 8677671 }  \\
\hline
  1853.25710        & 73.0   &    1.5       &  -155.6  &   3.1  \\
  1853.27307        & 73.7   &    1.7       &  -158.9  &  3.5  \\
  1853.28904        & 74.7   &    2.1       & -159.9   &  5.3   \\
  1866.18974        & -14.6  &    1.5       & 209.6    & 7.9   \\
  1866.20640        & -13.8  &    1.8       &  208.5   & 7.0     \\
  1866.22237        & -12.5  &    1.8       & 205.0    &  4.1     \\
\hline
\multicolumn{5}{c}{TIC 318217844}   \\
\hline
   1123.24526      &  6.6     &  3.7      &   -110.6    &   4.8  \\
   1123.25499      &  16.0    &  3.2      &   -112.5    &  5.2  \\
   1123.26401      & 22.0     &  3.4      &   -102.6    &  4.9  \\
   1482.27514      & -148.1   &  2.5      &    119.4    &  3.2   \\
   1482.29180      &  -143.0  &  2.8      &    103.9    & 3.9   \\
   1482.30778      &  -139.3  &  4.2      &    106.3    &  6.7   \\
   1482.32375      &  -133.6  &  2.2      &    101.7    & 3.8    \\
   1482.34041      &  -127.3  &  4.0      &    103.1    & 4.9   \\
   1482.35639      &  -121.3  &  3.6      &    90.2     & 5.2   \\
   1482.37236      &  -114.0  &  3.8      &    102.6    &  5.8   \\
   1798.39667      &  48.9    &  3.0      &    -118.8   &  4.7  \\
   2188.34320      & 16.1     &  4.1      &     -119.1  &  5.7 \\
   2188.35918      & 14.9     &  4.3      &    -101.5   &  6.2 \\
   2188.37516      & 2.9      &  5.6      &    -111.0   &  7.8  \\
   2188.39181      & 2.1      &  5.8      &    -87.7    &   8.3 \\
   2188.40778      &  14.1    &  5.5      &    -72.0    &   8.0  \\
   2188.42377      & -3.5     &  6.1      &    -3.7     &  8.2  \\
\hline
\end{tabular}
\footnotesize {}
\end{table}

\label{lastpage}

\end{document}